\documentclass[11pt,twoside,letterpaper]{article} %% The same for {book}
\usepackage{times,fancyhdr,ulem,amsmath,amssymb}
\usepackage[dvips]{graphicx}

\makeatletter
\def\cleardoublepage{\clearpage\if@twoside \ifodd\c@page\else%
    \hbox{}%
    \thispagestyle{empty}%
    \newpage%
    \if@twocolumn\hbox{}\newpage\fi\fi\fi}
\makeatother

\def\figurename{Figure}
\makeatletter
\renewcommand{\fnum@figure}[1]{\figurename~\thefigure.}
\makeatother

\def\tablename{Table}
\makeatletter
\renewcommand{\fnum@table}[1]{\tablename~\thetable.}
\makeatother
\begin{document}
\title{
{\begin{flushleft}
\vskip 0.45in
{\normalsize\bfseries\textit{Chapter~}}
\end{flushleft}
\vskip 0.45in
%
%
%%%%%%%%%%%%%%%%%%%%%%%%%%%%%%%%%%%%%%%%%%%%%%%%%%%%%%%%
%
%
% AUTHOR:  This belongs to you
%%%%%%%%%%%%%%%%%%
\bfseries\scshape  The inner crust and its structure}}
\author{\bfseries\itshape D\'ebora P. Menezes$^{(1)}$,  Sidney S. Avancini$^{(1)}$,\\ 
  \bfseries\itshape Constan\c ca Provid\^encia$^{(2)}$ and  \bfseries\itshape Marcelo D. Alloy$^{(3)}$
\thanks{E-mail debora@fsc.ufsc.br}\\
$^{(1)}$ Depto de F\'{\i}sica - Universidade Federal de Santa Catarina, \\CP 476 -
CEP 88040-900, Florian\'opolis, SC, Brazil\\
$^{(2)}$ Centro de F\'{\i}sica Computacional, Departamento de F\'{\i}sica,\\ Universidade de
Coimbra, 3004-516 Coimbra, Portugal\\
$^{(3)}$ Universidade Federal da Fronteira Sul,\\ Chapec\'o, SC, CEP 89.812-000,
Brazil\\}
\date{}
\maketitle
\thispagestyle{empty}
\setcounter{page}{1}
% ------- [First Page Running Head] - place it immediately after title! ------
%\thispagestyle{fancy}
%\fancyhead{}
%\fancyhead[L]{In: Neutron Star Crust \\
%Editors: C.A. Bertulani and J. Piekarewicz, pp. {\thepage-\pageref{lastpage-01}}} % needs \label{lastpage-01} on the last page.
%\fancyhead[R]{ISBN 0000000000  \\
%\copyright~2012 Nova Science Publishers, Inc.}
%\fancyfoot{}
%\renewcommand{\headrulewidth}{0pt}
%
%\vspace{2in}
%
%
% ------------ [Running Heads - for odd and even pages] - please insert it only on page 2!
\pagestyle{fancy}
\fancyhead{}
\fancyhead[EC]{D.P. Menezes,  S.S. Avancini,  C. Provid\^encia  and M.D. Alloy}
\fancyhead[EL,OR]{\thepage}
\fancyhead[OC]{The inner crust and its structure}
\fancyfoot{}
\renewcommand\headrulewidth{0.5pt}
%------------------------------------------------------------------------------
%
\begin{abstract}
In this chapter we discuss some possible physical pictures that
describe the constitution of the inner crust of compact objects.
Different relativistic models both with constant couplings and
density dependent ones are used. We  calculate the liquid-gas phase
transition in asymmetric nuclear matter from the  thermodynamic and dynamic
instabilities. The equations of state used to describe the crust are
related to the crust-core transition properties.
Cold and warm pasta phases with and without alpha particles
are constructed. The influence of the pasta phase and its internal
structure on the diffusion coefficients associated with Boltzman transport
equations used to simulate the evolution of protoneutron stars are shown.
Finally, the possible existence of bare quark stars and the effects of
strong magnetic fields on quark matter are considered.
Open questions are pointed out.
\end{abstract}

\par

\noindent \textbf{PACS} 05.45-a, 52.35.Mw, 96.50.Fm.

\noindent \textbf{Keywords:} phase transitions, pasta phase,
quark stars

\section{Introduction}

The internal constituents of compact stars are a great source of speculation
\cite{glen}.
They could be made of hadronic matter only, quark matter only (a possibility
arisen by the Bodmer-Witten conjecture \cite{bodmer,olinto}), or they could be
hybrid. Hybrid stars may have in their interior hadrons and quarks with or
without a condensate. If the star is composed of quark matter only,
it may be a bare star and we will tackle this point at the end of this
chapter. On the other hand, if it is made of hadrons only or if it is a
hybrid object, it has a crust, our main interest in what follows.
The crust mass and thickness depend on the equation of state (EOS) that
  describes the star. They also depend on the total stellar mass; an increase of the
  stellar mass increases the gravitational pull within the crust
  resulting in the thinning of the crust. An estimation of the size of
  the crust done in \cite{haensel} gives values varying from 1.01 to
  0.29 Km.
 In particular, we will discuss the possible existence of inhomogeneous
structures, i. e. the pasta phase, in  the inner crust due to the competition between the
strong and Coulomb interactions and their implications for the properties of the crust.

Through out this chapter, we consider a system of protons and neutrons with
mass $M$
interacting with and through an isoscalar-scalar field $\phi$ with mass
$m_s$,  an isoscalar-vector field $V^{\mu}$ with mass
$m_v$, an isovector-vector field  $\mathbf b^{\mu}$ with mass
$m_\rho$.  We also include a system of leptons composed by electrons and
muons, electrons and neutrinos or just electrons, depending on the problem. The
Lagrangian density reads:
\begin{equation}
\mathcal{L}=\sum_{i=p,n}\mathcal{L}_{i}\mathcal{\,+L}_{{\sigma }}%
\mathcal{+L}_{{\omega }}\mathcal{+L}_{{\rho }}\mathcal{+L}_{{\gamma }}
+\sum_{l=e,\nu, \mu} \mathcal{L}_l,
\label{lag}
\end{equation}
where the nucleon Lagrangian density has the form
\begin{equation}
\mathcal{L}_{i}=\bar{\psi}_{i}\left[ \gamma _{\mu }iD^{\mu }-M^{*}\right]
\psi _{i}  \label{lagnucl},
\end{equation}
with
\begin{eqnarray}
iD^{\mu } &=&i\partial ^{\mu }-\Gamma_{v}V^{\mu }-\frac{\Gamma_{\rho }}{2}
{\vec{\tau}}%
\cdot \mathbf{b}^{\mu } - e \frac{1+\tau_3}{2}A^{\mu}, \label{Dmu} \\
M^{*} &=&M-\Gamma_{s}\phi.
\label{Mstar}
\end{eqnarray}
The lepton Lagrangian density is given by
\begin{equation}
\mathcal{L}_l=\bar \psi_l\left[\gamma_\mu\left(i\partial^{\mu}+ e A^{\mu}\right)
-m_l\right]\psi_l,
\label{lage}
\end{equation}
where $e$, $m_l$ stand for the charge and mass of the lepton, respectively,
and the meson Lagrangian densities are
\begin{eqnarray*}
\mathcal{L}_{{\sigma }} &=&\frac{1}{2}\left( \partial _{\mu }\phi \partial %
^{\mu }\phi -m_{s}^{2}\phi ^{2}-\frac{1}{3!}\kappa \phi ^{3}-\frac{1}{4!}%
\lambda \phi ^{4}\right)  ,\\
\mathcal{L}_{{\omega }} &=&\frac{1}{2} \left(-\frac{1}{2} \Omega _{\mu \nu }
\Omega ^{\mu \nu }+ m_{v}^{2}V_{\mu }V^{\mu } \right) , \\
\mathcal{L}_{{\rho }} &=&\frac{1}{2} \left(-\frac{1}{2}
\mathbf{B}_{\mu \nu }\cdot \mathbf{B}^{\mu
\nu }+ m_{\rho }^{2}\mathbf{b}_{\mu }\cdot \mathbf{b}^{\mu } \right) ,\\
\mathcal{L}_{{\gamma }} &=&-\frac{1}{4}F _{\mu \nu }F^{\mu
  \nu },
\end{eqnarray*}
where $\Omega _{\mu \nu }=\partial _{\mu }V_{\nu }-\partial _{\nu }V_{\mu }$
, $\mathbf{B}_{\mu \nu }=\partial _{\mu }\mathbf{b}_{\nu }-\partial _{\nu }
\mathbf{b}_{\mu }-\Gamma_{\rho }(\mathbf{b}_{\mu }\times \mathbf{b}_{\nu })$
and $F_{\mu \nu }=\partial _{\mu }A_{\nu }-\partial _{\nu }A_{\mu }$.
The  parameters of the models  are the nucleon mass $M=939$ MeV,
the coupling parameters $\Gamma_s$, $\Gamma_v$, $\Gamma_{\rho}$ of the mesons to
the nucleons, the electron mass $m_e=0.511$ MeV, the muon mass $m_\mu=105.66$
MeV and the electromagnetic
coupling constant $e=\sqrt{4 \pi/137}$. The electron neutrino mass is considered to be zero.
In the above Lagrangian density $\vec \tau$ are the Pauli matrices. When density dependent
models are used,
the non-linear terms are not present and, hence, $\kappa=\lambda=0$ and the
density dependent parameters are chosen as in \cite{tw,gaitanos,inst04}. For
the parametrizations with constant couplings, $\Gamma_i$ is replaced
by $g_i$, where $i=s,v,\rho$ as in the NL3 parameter set \cite{nl3}.

Some expressions are often used next. $Y_p$ and $Y_L$ refer to the
proton and lepton fractions respectively. If matter in $\beta$-equilibrium is considered in
a system of protons, neutrons, electrons and possibly trapped electron neutrinos, one has:
\begin{equation}
\mu_p=\mu_n-(\mu_e -\mu_\nu).
\end{equation}
For neutrino free matter $\mu_\nu=0$.
Neutrality of charge requires $\rho_p=\rho_e + \rho_{\mu}$ and, if
$\rho_\mu$ is present, its chemical potential is equal to the electron
chemical potential.

\section{Instabilities in Nuclear Matter}

It was shown in \cite{pethick95,link} that the pressure and density at the
inner boundary of the crust
(transition pressure and transition density)
define the mass and moment of
inertia of the crust. This establishes a relation between
the equation of state (EOS) and compact-star observables. In this section we show how an
estimation of the transition properties may be obtained from the thermodynamical binodal and
spinodal surfaces, or the dynamical spinodal. A better estimation, obtained from the pasta
phase calculation, will be discussed in section \ref{pasta}

The liquid-gas phase transition at subsaturation densities  is a well known feature of the nuclear EOS \cite{muller95}. It corresponds to the presence of a negative curvature of the free energy density $\cal F$. In this case
the system is unstable against separation into two infinite homogeneous phases. The spinodal surface, which limits the unstable region in the $(T,\rho_p,\rho_n)$ space, is defined by the cancellation of the determinant of the free energy curvature matrix \cite{avancini06}:
 $$  {\cal C}_{ij}=\left(\frac{\partial^2{\cal F}}{\partial \rho_i\partial\rho_j}
 \right)_T,
\qquad   \mathcal{C=}\left(
     \begin{array}{cc}
\frac{\partial \mu _{n}}{\partial \rho _{n}} & \frac{\partial \mu _{n}}{%
\partial \rho _{p}}  \\
\frac{\partial \mu _{p}}{\partial \rho _{n}} & \frac{\partial \mu _{p}}{%
\partial \rho _{p}}
\end{array}
\right).$$

 The eigenvalues of the curvature matrix are given by
  $$  \lambda_{\pm}=\frac{1}{2}
\left(\mbox{Tr}({\cal C})\pm\sqrt{\mbox{Tr}({\cal C})^2-4\mbox{Det}({\cal C)}}\right),
    $$
and are associated to the  eigenvectors $\boldsymbol{\delta\rho_\pm}$.
Inside the spinodal region the lowest eigenvalue $\lambda_-$ is negative. The direction of the
vector $\boldsymbol{\delta\rho_-}$ defines the  direction of instability, which generally
dictates a distillation effect corresponding to a phase separation into a high density
symmetric matter and low density neutron rich matter \cite{avancini06}.

\begin{figure}[hbt]
\begin{center}
   \begin{tabular}{cc}
\includegraphics[width=0.56\linewidth]{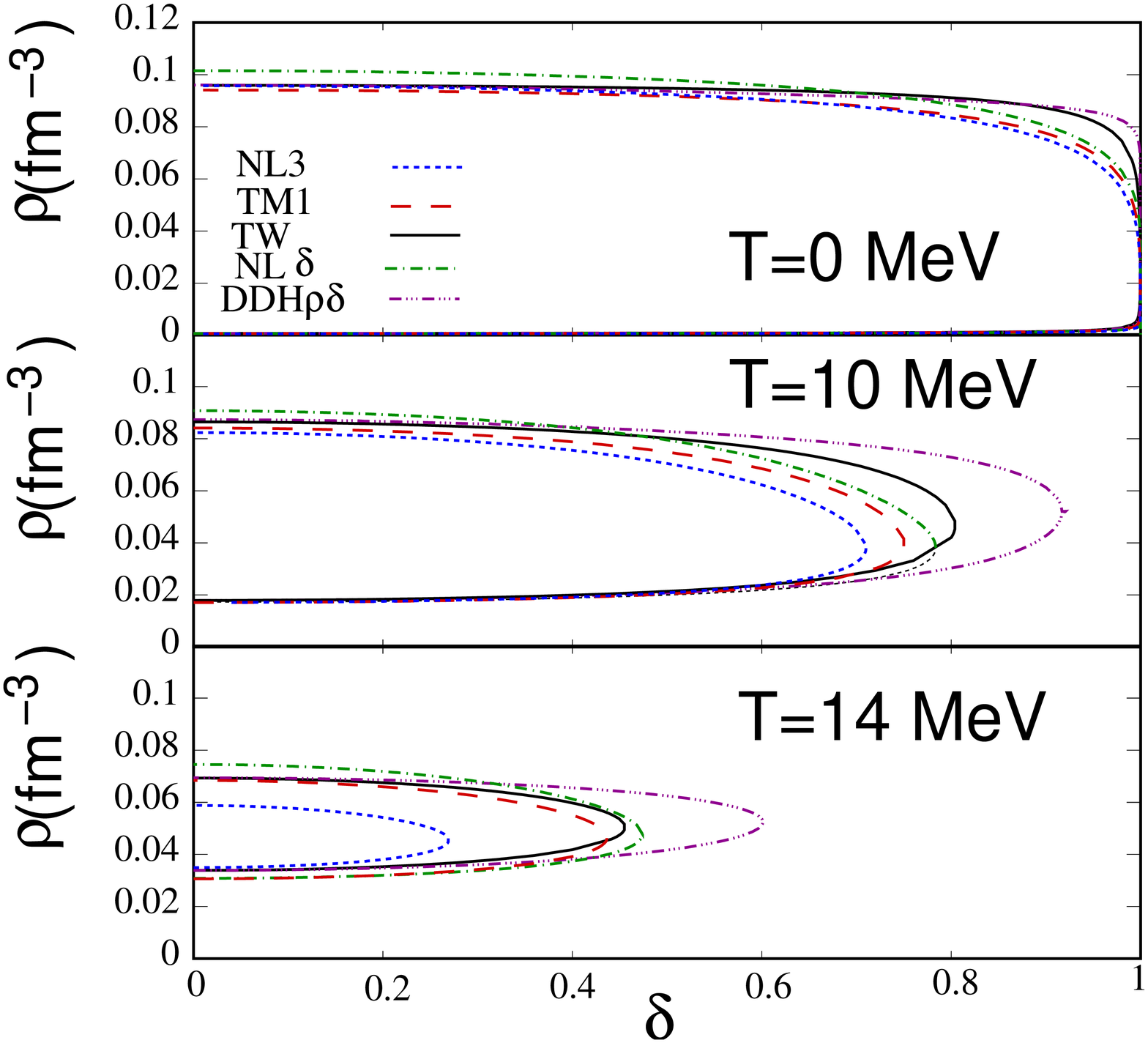}&
\includegraphics[width=0.35\linewidth]{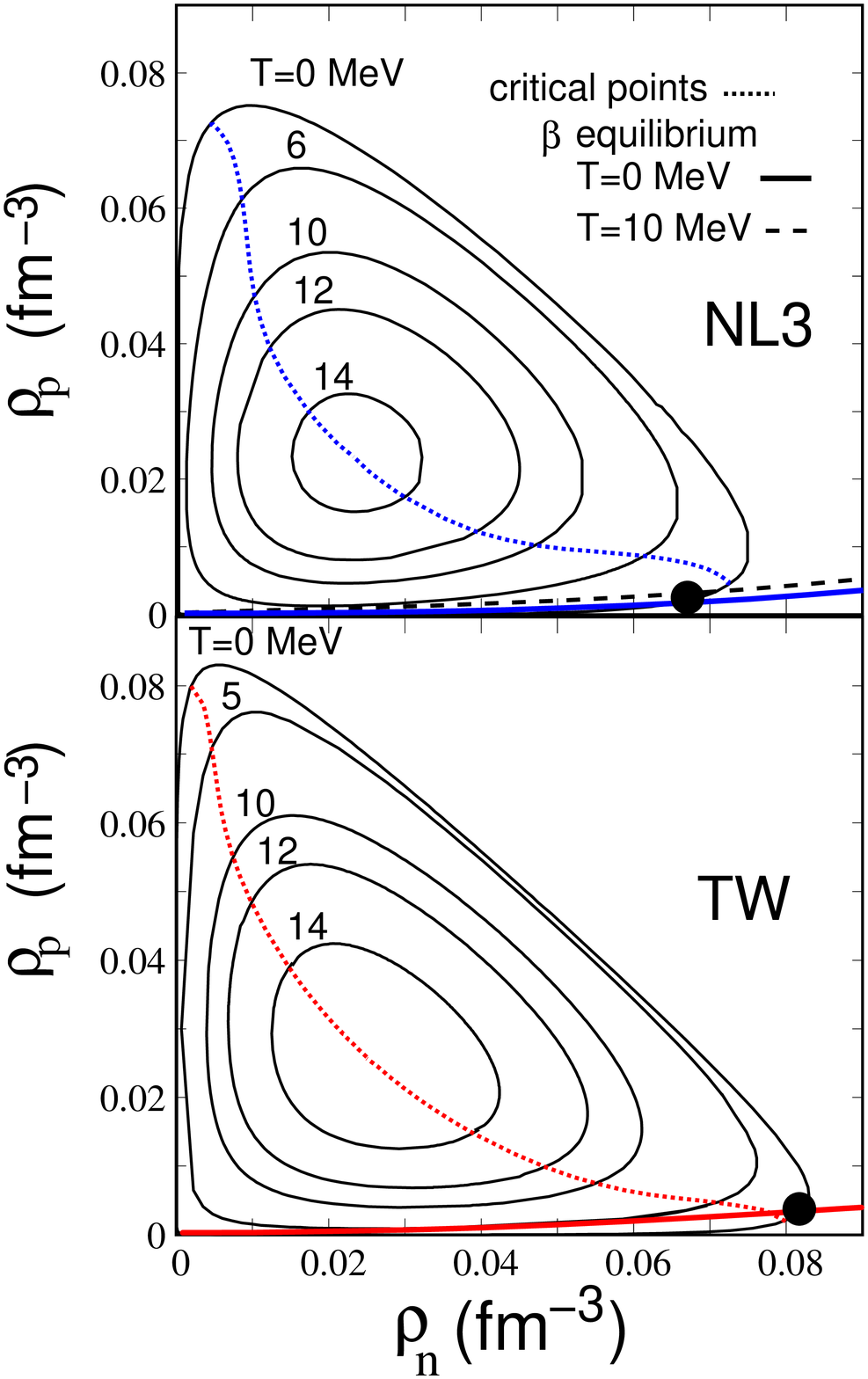}
  \end{tabular}
  \caption{Left panel: the spinodal section for different RMF
    models models in the $\rho-\delta$ plane. Fig. taken from \cite{inst04}.
Right panel: the spinodal section for NL3 and TW and different temperatures. The
$\beta$-equilibrium EOS at $T=0$ ($T=10$) MeV is represented by a
full (dashed) line. The full dot identifies the transition crust-core
at $T=0$. Fig. taken from \cite{avancini06}.}
\label{inst}
\end{center}
\end{figure}

In the left panel of Fig.\ref{inst} the spinodal sections obtained
for different relativistic mean field
(RMF) models are plotted in terms of the total density $\rho$ and
the asymmetry parameter
\begin{equation}
\delta=(\rho_n-\rho_p)/\rho = 1 - 2 Y_p.
\label{delta}
\end{equation}
While for symmetric matter most of the models show a similar behavior, at high asymmetry
and/or temperature models differ. For this reason, establishing
constraints on the equations of state based on experimental results
is an important step, and a discussion on this aspect is done in
section \ref{vinculos}
In the right panel of Fig. \ref{inst}, the spinodal sections for several temperatures and two models
(NL3 \cite{nl3} and TW \cite{tw}) are an
example of the possible existing differences: for TW with a symmetry energy and corresponding
slope equal, respectively, to 32.76 and 55.30 MeV, the transition density (identified by a full
dot) occurs at a larger density and for a smaller proton fraction than for NL3 with a symmetry
energy and slope equal, respectively, to 37.34 and 118.30 MeV. Also, for TW the spinodal
surface extends to a larger temperature.

Instead of using a thermodynamic  approach, in \cite{pethick95a}  the transition density  was
estimated in  a  local equilibrium approximation by calculating the density for which  matter
becomes unstable to  small density fluctuations. An equivalent approach is the Vlasov
formalism, a semi-classical limit of the description of the  collisionless regime, which has
been used in \cite{brito06} to calculate the dynamical spinodal within several RMF
models. Contrary to the calculation of the thermodynamical spinodal, the effect of the finite
range of the nuclear interaction as well as the Coulomb interaction and the presence of
electrons is taken into account in the determination of the dynamical spinodal.

In the Vlasov formalism  the equilibrium state, characterized by the Fermi momenta of neutrons,
protons and electrons,   $p_{Fn}, \,p_{Fp},\, p_{Fe}$, is the starting point.
Charge neutrality requires $p_{Fe}=p_{Fp}$. A perturbation of the system is then described by
the perturbed mesonic fields, $F_i\,=\, F_{i0} + \delta F_i\;,$ and the perturbed distribution
functions for the neutrons, protons and electrons
    $$
    f_i(t,\boldsymbol{r,\, p}\,)=\, f_{0i} + \delta f_i\;, \quad
    \delta f_i \,=\, \{S_i,f_{0i}\},$$
where {\bf $f_{0i}=\theta (p_{F_i}^2-p^2)$ }and the generating function $S({\bf r},{\bf p},t)=\mbox{diag}\left(
      S_p, \, S_n,\, S_e
    \right)$ has been introduced.
The time evolution of the distribution function $f_i$ is described by the Vlasov equation \cite{brito06}
    $$
    \frac{\partial f_i}{\partial t} +\{f_i,h_i\}=0, \qquad \; i=p,\,n,\,e.
    $$
In the limit of small perturbations  the linearized Vlasov equation
    $$\frac{d{\cal S}_{i}}{d t}+ \{{\cal S}_{i}, h_{0i}\}    =\delta h_{i}$$
is solved. In this equation  $h_{0i}$ and $\delta h_{i}$ stand, respectively, for the
equilibrium single particle Hamiltonian  and corresponding perturbation from  equilibrium,
and $\{v,w\}$ represents the Poisson bracket of two dynamical functions $v$ and $w$.
 Longitudinal fluctuations are described by the ansatz
    $$
    \left(
      \begin{array}{llll}S_i &   \delta F_j & \delta \rho_i & \delta h_i \end{array} \right)
    =
    \left(
      \begin{array}{llll}S_{\omega,i}(x) & \delta F_{\omega,j} &\delta \rho_{\omega,i} &        \delta h_{\omega,i} \end{array}\right)
    e^{i(\boldsymbol{q\cdot r}-\omega t)} ,
$$
where $\boldsymbol q$ and $\omega$ are the transferred momentum and energy  and
$x=\cos(\boldsymbol{p\cdot q})$. The dynamical spinodal surface is characterized by a zero
frequency $\omega$.

\begin{figure}[hbt]
\begin{center}
\includegraphics[width=0.9\linewidth]{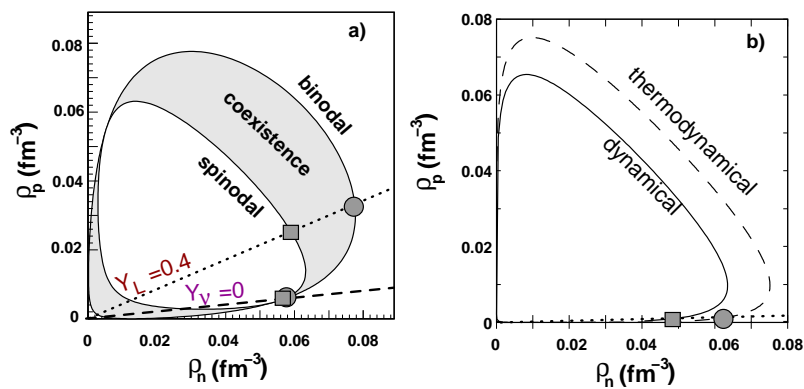}
  \caption{The transition density for neutrino free $\beta$-equilibrium matter
    and matter with trapped neutrinos and a fraction of leptons $Y_L=0.4$: a) estimation from
the spinodal and  binodal sections; b) estimation from the dynamical and thermodynamical
spinodal.}
\label{binodal}
\end{center}
\end{figure}

It is expected that the transition density lies in the metastable region
between the binodal surface and the dynamical spinodal surface.
 The binodal surface is defined  in the $\rho,\, Y_p,\, T$ phase space
as the surface where the gas and liquid phases coexist, and it defines an  upper limit for the
extension of the pasta phase since it also does not take into account Coulomb nor finite
size effects. The binodal surface is calculated imposing Gibbs conditions: for a given
temperature,  the pressure and
the proton and neutron chemical potentials are equal in both phases \cite{muller95}.
 The thermodynamical
spinodal touches the binodal surface at the critical point, which, for a given temperature,
occurs for the largest pressure on both surfaces and at a density and
proton fraction close to the crust-core transition density and  proton fraction of
cold stellar $\beta$-equilibrium matter. This is represented in Fig. \ref{binodal}a) where the
square and the circle represent respectively the crust-core transition from the thermodynamical
spinodal and binodal. Both the EOS of $\beta$-equilibrium neutrino free matter and matter with
trapped neutrinos are represented. We conclude that for neutrino free matter both
estimations almost coincide, while for matter with a lepton fraction $Y_L$=0.4 there is a large
difference. As stated before,  it is possible to get a better estimation of the lower
limit of the pasta phase extension if, instead of the thermodynamical spinodal, the dynamical
spinodal is calculated, as in Fig. \ref{binodal}b). Taking into account finite range effects and
electrons makes the spinodal region smaller, so even though the thermodynamical method gives a
good estimation  of the transition density for cold
$\beta$-equilibrium matter, it is a bit too large. This is confirmed by pasta calculations
\cite{warmpasta} as seen in the next section.

For densities inside the dynamical spinodal matter is non-homogeneous, and for densities outside
the binodal surface, matter is homogeneous. Between these two surfaces we may find matter in
a metastable configuration. The most probable configuration is the one with the smallest free
energy density and requires a pasta phase calculation, the topic of the next section.

\section{Cold and Warm Pasta Phase}\label{pasta}

In the inner crust of a neutron star (NS), a liquid gas phase transition can give rise to the
existence of the pasta phase, which is a frustrated system that arises from the
competition between the strong and the electromagnetic interactions. The pasta
phase appears at densities on the order of 0.001 - 0.1 fm$^{-3}$
in neutral nuclear matter and at a smaller density range in $\beta$-equilibrium
 stellar matter. The basic shapes of these structures were named
as droplets (bubbles), rods (tubes) and slabs for three, two and one dimensions
respectively. The ground-state configuration is the one that minimizes the
free energy. In what follows we use two different prescriptions in order to
build the pasta phase at zero and finite temperatures: one is based on phase
coexistence (CP) and obeys the Gibbs conditions, and the other is the
Thomas-Fermi (TF) approximation.
 Within the CP method, the pasta structures are built with different
geometrical forms in a background nucleon gas. This is achieved by calculating
from the Gibbs' conditions the density and the proton fraction of the pasta and
of the background gas, so that in the whole we have to
 solve simultaneously  various equations. These equations are related to the fact that the
 pressure, proton and neutron chemical potentials and temperature are
 the same in both phases. Two more equations are
 related to the nucleon effective mass in each phase, and an equation
 that balances the amount of protons in each phase has also to be solved:
\begin{equation}
f (\rho_p^I) + (1-f) (\rho_p^{II}) = Y_p \rho,\label{gibbs1}
\end{equation}
where I and II label each of the phases, $f$ is the volume fraction of
phase I:
\begin{equation}
f= \frac{\rho -\rho^{II}}{\rho^I-\rho^{II}}, \label{volf}
\end{equation}
where the total baryonic density is
\begin{equation}
\rho=\rho_p + \rho_n,
\end{equation}
and $Y_p$ is the global proton fraction.

The correct parametrization of the surface energy, which is temperature, proton fraction
and geometry dependent, must be used \cite{pasta_alpha}.  The
following functional for the surface tension coefficient, $\sigma$, is used,
\begin{equation}
\sigma = \sigma (x,T=0)\left[ 1-a(T)~x \rm{T} -b(T)T^2 \right]~, \label{sigpar}
\end{equation}
where  $x=\delta^2$  stands for the square of the asymmetry parameter defined in eq.(\ref{delta}).
The CP method is very easy to implement;
however,  since it is not self-consistent, the results obtained within the method  should be
taken with care. The self-consistent Thomas Fermi results should be compared with other
more realistic, yet more involving methods that do not impose a pre-defined form, such as
the  quantum molecular dynamics calculation in \cite{watanabe08} or the 3D
Skyrme-Hartree-Fock method at finite temperature used in \cite{newton09}.

\begin{figure}[hbt]
\begin{center}
\includegraphics[width=1.0\linewidth]{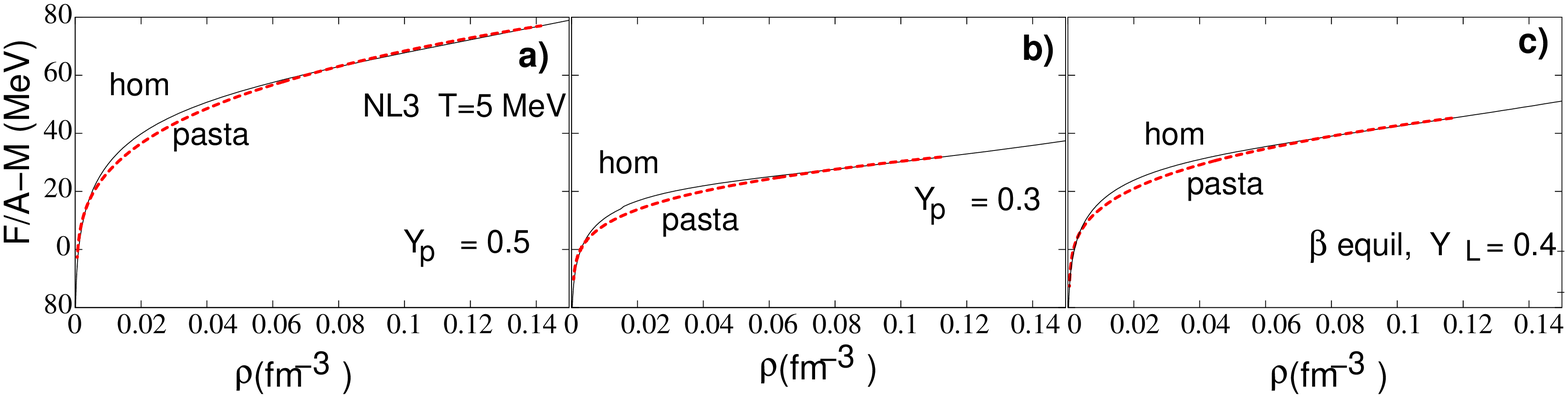} \\
\end{center}
\caption{Free energy for the homogeneous and pasta like matter.
The red dashed line defines the pasta free energy. The crossing of the pasta
and the homogeneous free energy define the extension of the pasta phase.}
\label{free}
\end{figure}

In Fig. \ref{free} we plot the free energy per particle for homogeneous and pasta like
matter described by the NL3 model within CP, with T=5 MeV and
proton fractions $Y_p=0.5,\,0.3$ and for $\beta$-equilibrium matter with trapped
neutrinos for a lepton fraction $Y_L=0.4$. This figure illustrates well the effect of the pasta
phase on the free energy: the formation of a non-homogeneous phase is energetically
favored.  The system in equilibrium chooses the configuration with the
lowest free energy, so the pasta-like matter defines the ground state of the
system  if its free energy is lower than the free energy of the
corresponding homogeneous matter.
The upper density limit of the pasta phase lies inside the binodal surface and decreases when
the proton fraction decreases. For $\beta$-equilibrium matter this limit defines the NS
crust-core transition.

\begin{figure}[hbt]
\begin{center}
\begin{tabular}{cc}
\includegraphics[width=0.4\linewidth]{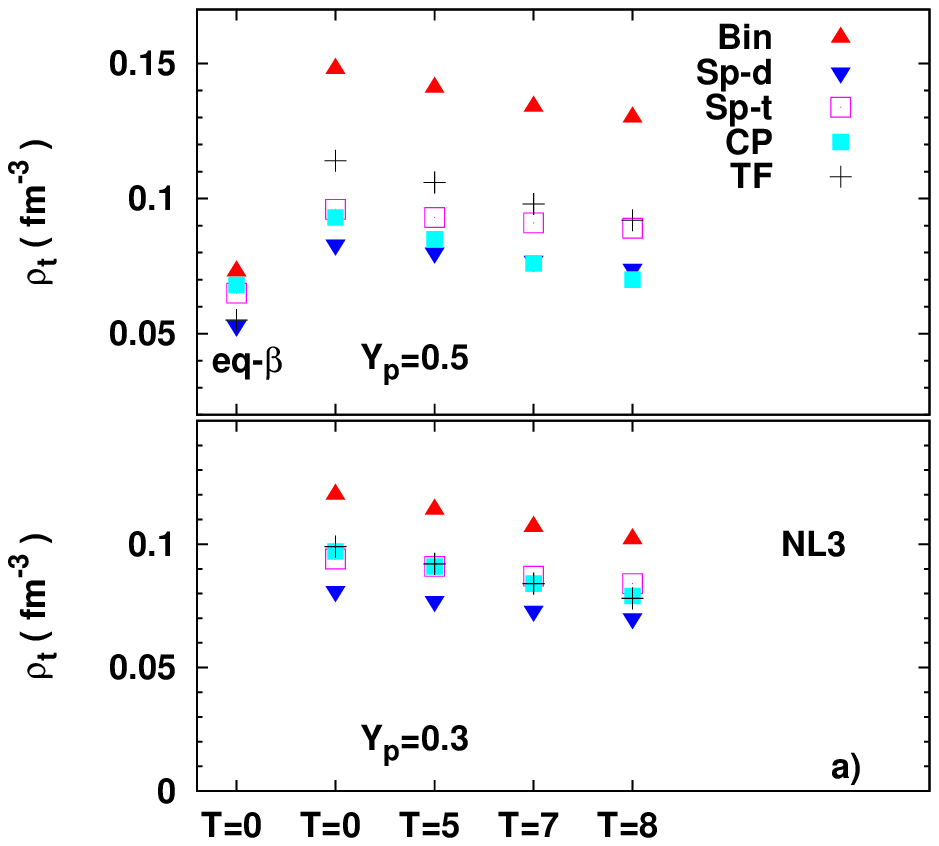} &
\includegraphics[width=0.4\linewidth]{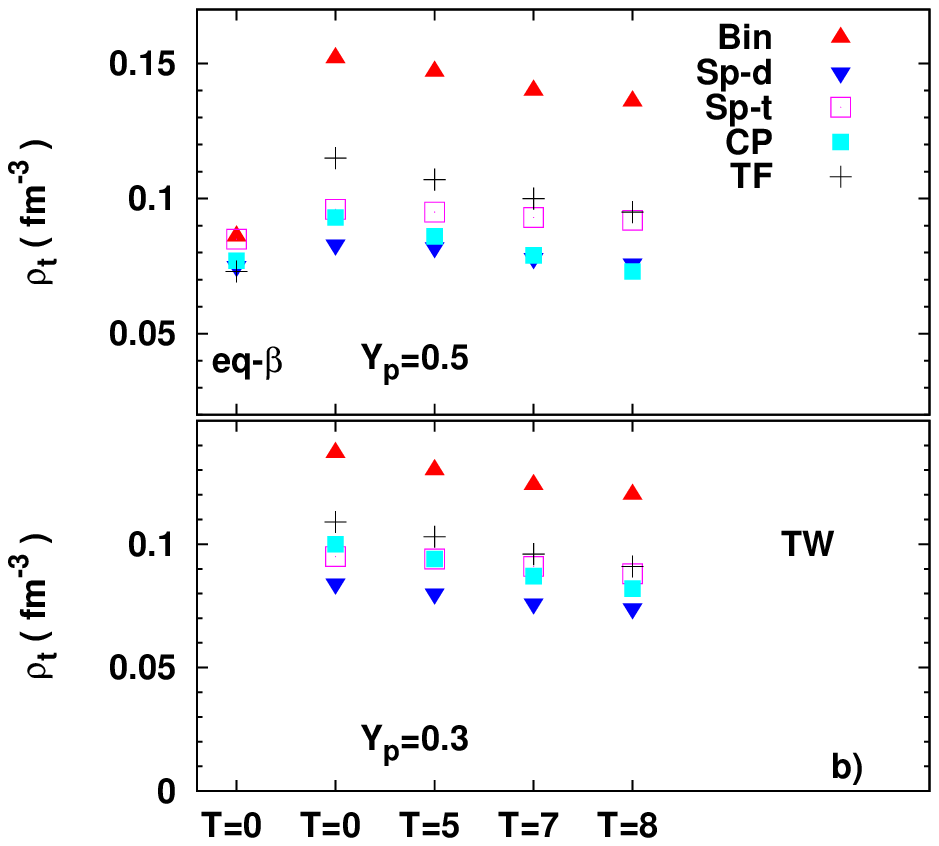} \\
\end{tabular}
\end{center}
\caption{Transition density, for several temperatures
and proton fractions $Y_p=0.5$ (upper plots), $0.3$ (lower plots)
and $\beta$-equilibrium (eq-$\beta$) at T=0 (left part of upper plots),
obtained using different methods within the models a) NL3 and b) TW. 
Fig. taken from \cite{erratum}.}
\label{figtrans}
\end{figure}

In Fig. \ref{figtrans} we compare the  estimations for transition density from a
non-homogeneous phase to a homogeneous phase obtained from the binodal surface (Bin)
\cite{muller95}, the dynamical spinodal surface (Sp-d) \cite{pethick95a,brito06}
the thermodynamic spinodal surface (Sp-t) \cite{avancini06}
and from the CP and TF methods. The TF results lie always between
the results obtained from  the dynamical spinodal and the binodal surfaces.
This is a self-consistent method that should satisfy these two constraints.
The CP calculation may predict results at densities  that lie below the dynamical spinodal
 value, namely at high temperatures and symmetric matter. For CP to
 give reasonable results, it is important that a good parametrization
 of the surface energy is used. The thermodynamical-spinodal calculation always  gives a quite good
estimation even though it does not take into account neither the surface nor
the Coulomb effects. For $\beta$-equilibrium matter all methods give
similar results. This is due to the occurrence of the transition density close
to the critical point where the spinodal and binodal surfaces touch and the
pressure on these surfaces is maximum.

\begin{figure}[hbt]
\begin{center}
\begin{tabular}{cc}
\includegraphics[width=0.4\linewidth]{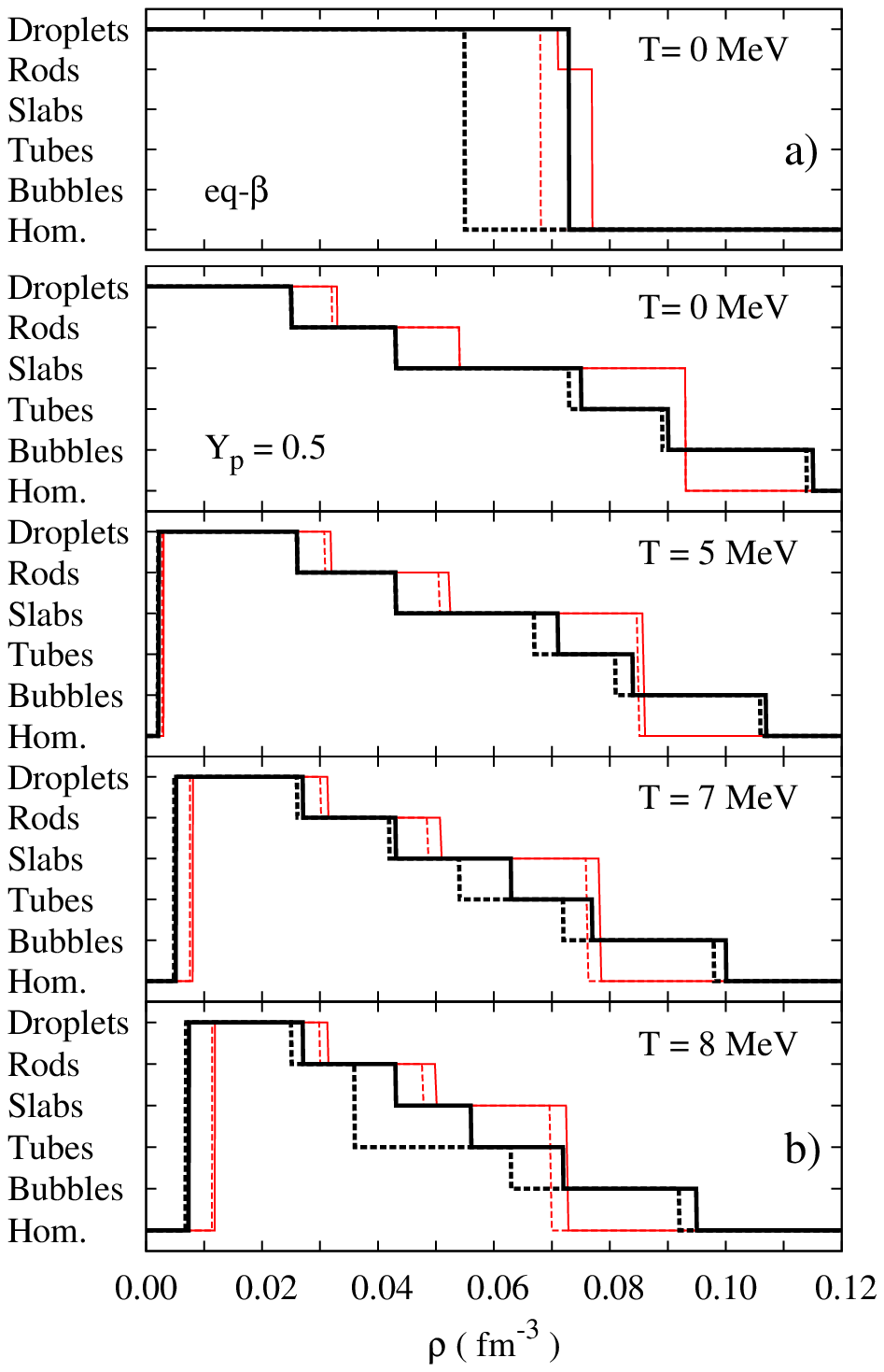} &
\includegraphics[width=0.4\linewidth]{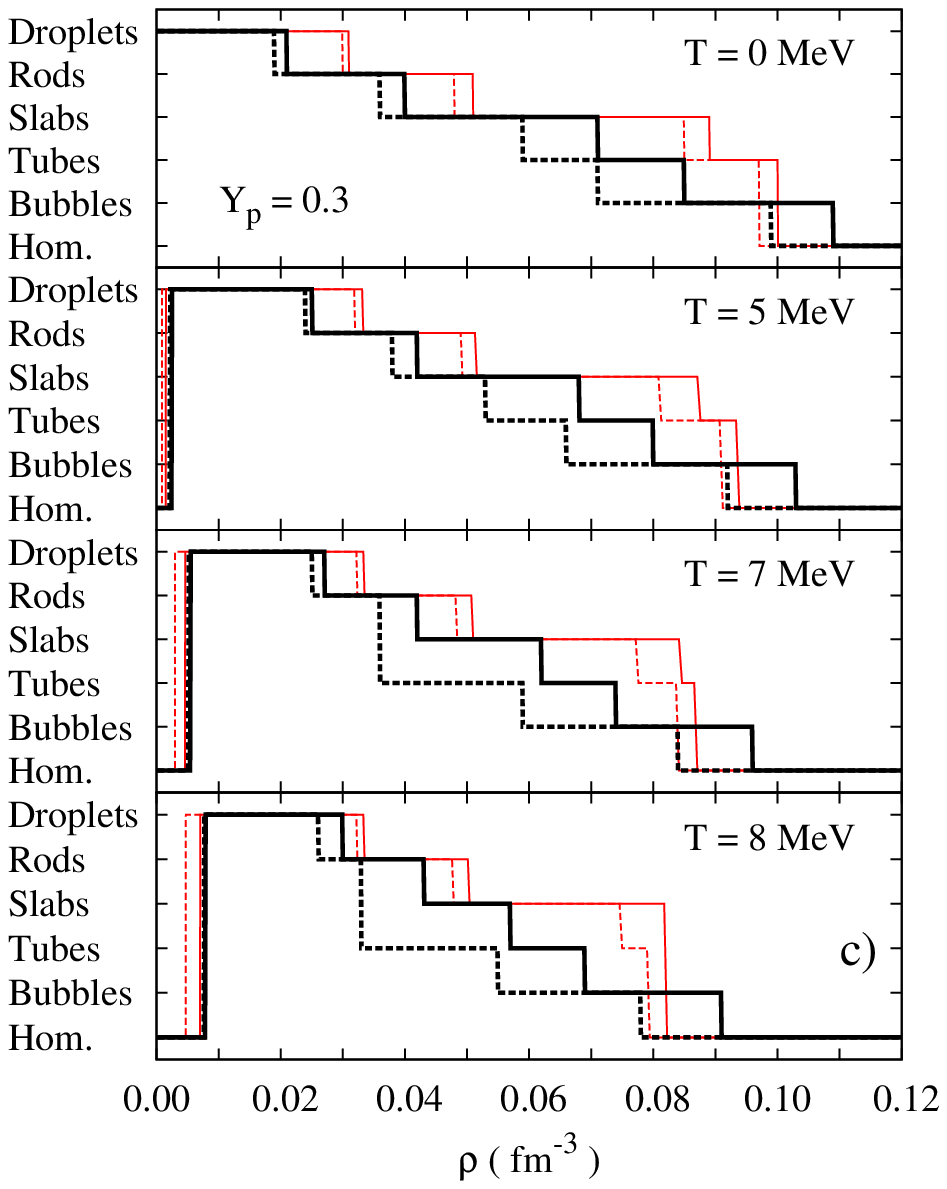} \\
\end{tabular}
\end{center}
\caption{Pasta phases: comparison between  NL3 (dashed line)  and TW (full line)
and the methods CP (thin red lines) and TF (thick lines) for
a) $\beta$-equilibrium cold stellar matter;
b) $Y_p=0.5$, c) $Y_p=0.3$. Fig. taken from \cite{erratum}.}
\label{figpasta}
\end{figure}

In Fig. \ref{figpasta} we compare the  density range for which each pasta
configuration exists within NL3 and TW. The thick lines stand for the
Thomas-Fermi calculation, the thin ones for the CP method. Full lines
represent TW and dashed ones NL3. For symmetric matter the main difference between the models
is the appearance of the different phases at slightly smaller densities
within NL3. However,  the largest differences occur for $Y_p=0.3$: NL3 has a much smaller rod
phase and  no slab phase
at T=7 and 8 MeV and quite large tube and bubble phases. The figure also illustrates the power
and limitations of the simpler method CP: the onset at lower densities of the pasta phase is
quite well described, however, it fails to describe the bubble phase and predicts a
smaller crust-core transition density.

It is important to point out that Figs. \ref{figtrans} and
\ref{figpasta} are slightly different from the ones published in
\cite{warmpasta} for the cases $Y_p=0.3$ and matter in
$\beta$-equilibrium because a slightly different paramerization of the
surface coefficient was used, i.e., in eq.(\ref{delta}), the proton
fraction was taken as the proton fraction of the denser phase in
\cite{warmpasta} and here it was taken as the global proton fraction.

The importance of the $\alpha$ particles cannot be neglected. It is the most
strongly bound system among all light clusters and it certainly plays a role in
nuclear matter, mainly at low temperatures. The Lagrangian density that describes the $\alpha$ particles
can be added to  Eq. \ref{lag} and is given by \cite{blaschke09}:
\begin{equation}
\mathcal{L}_{\alpha }=\frac{1}{2} (i D^{\mu}_{\alpha} \phi_{\alpha})^*
(i D_{\mu \alpha} \phi_{\alpha})-\frac{1}{2}\phi_{\alpha}^* M_{\alpha}^2
\phi_{\alpha},
\end{equation}
with
$iD^\mu_\alpha = i \partial ^{\mu }_\alpha-\Gamma_{\alpha} V^{\mu },$
where
$M_{\alpha}= 4 M - B_{\alpha}, \quad B_{\alpha}=28.3 ~~{\rm MeV}.$

The coupling of the $\omega$ meson to the $\alpha$-particles
is included for mimicking
the excluded volume effect and, consequently, the $\alpha$ particles dissolution at
high densities.  The dissolution
density  obtained using the ansatz $\Gamma_\alpha=4 \Gamma_v$ is in agreement
with the dissolution densities obtained within a quantum statistical approach
\cite{blaschke09}. More careful studies are necessary in order to determine
the adequate meson-cluster couplings. When $\alpha$ particles are
included, Eq.(\ref{gibbs1}) becomes
\begin{equation}
f (\rho_p^I + 2 \rho_{\alpha}^I) + (1-f) (\rho_p^{II} + 2 \rho_{\alpha}^{II})
= Y_p \rho,\label{gibbs7}
\end{equation}
where I and II label each of the phases, $f$ is the volume fraction of
phase I given in eq.(\ref{volf}), where the total baryonic density is now
\begin{equation}
\rho=\rho_p + \rho_n + 4 \rho_{\alpha},
\end{equation}
and $Y_p$ is the global proton fraction given by
\begin{equation}
Y_p = \frac{\rho_p + 2 \rho_{\alpha}}{\rho}.
\end{equation}

\begin{figure}[hbt]
\begin{center}
\includegraphics[width=0.9\linewidth]{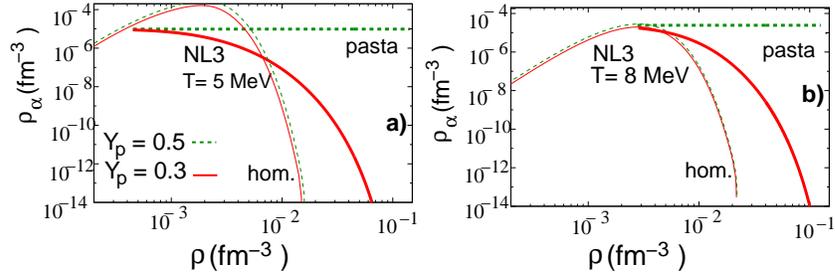} \\
\end{center}
\caption{$\alpha$ particle density for a)  T=5 MeV and b) T= 8 MeV and
$Y_p=0.3$ and 0.5 obtained with  NL3 for homogeneous
  matter (thin lines) and gas phase of the pasta-like matter (thick lines).
Fig. taken from \cite{pasta_alpha}.}
\label{figalpha}
\end{figure}

The $\alpha$ particle densities are plotted in Fig. \ref{figalpha}
for $Y_p=0.5$ and 0.3 and $T=5$  and 8 MeV. We include
the  calculation for both  homogeneous matter  and pasta-like matter.
This figure gives a hint on the possible effects of $\alpha$ particles in the
inner crust of a compact star, which is larger for the larger temperatures
and larger proton fractions. Due to the existence of a non-homogeneous
phase, the $\alpha$-particles dissolve at larger densities, although
the  $\alpha$ fraction may take very small values. It is important to include other small
clusters (deuteron, tritium and helium 3), which, due to their smaller masses will predominate
over the $\alpha$ particles at the larger temperatures. We expect that the appearance of these
light clusters will affect heat and transport properties.

The properties of the pasta phase are also discussed in the chapter
{\it Nuclear pasta in supernovae and neutron stars} by G. Watanabe and T.
Maruyama, where different formalism and numerical methods are used.

\section{Constraints on the Equations of State} \label{vinculos}

We have seen that the EOS that describe equally well the properties of nuclear saturation matter
and the ground-state properties of nuclei, predict quite  different spinodal surfaces at high
asymmetries and/or temperatures in Fig \ref{inst}. The same occurs for other properties at high densities, such as
the incompressibility.

 The asymmetric nuclear matter EOS may  be constrained by 
 the properties of asymmetric nuclear matter obtained from
various analyses of experimental data, including isospin diffusion measurements \cite{li08},
giant resonances \cite{garg07}, isobaric analog states \cite{danie09} or
meson production (pions \cite{li05}, kaons \cite{fuchs06}) in heavy
ion collisions (see \cite{yennello10} for an overview).

Some correlations between finite nuclei properties and bulk matter properties have been obtained:
 a linear relation between the density derivative of the neutron
matter EOS at 0.1 fm$^{-3}$ and the neutron skin thickness of $^{208}$Pb that  has been theoretically tested with different
Skyrme parameter sets \cite{brown00} and relativistic Hartree models \cite{typel01}. Another
linear correlation was obtained between the
$^{208}$Pb skin thickness and the liquid-to-solid transition density in neutron stars
\cite{horo01}.
In what follows the neutron skin thickness is defined as
\begin{equation}
\delta R=R_n-R_p, \label{skin}
\end{equation}
where the mean square radius is
\begin{equation}
R_{i}^{2}=\frac{\int~d^{3}r r^2 \rho_{i}(\mathbf {r})}{\int~d^{3}r
 \rho_{i}(\mathbf {r})}. \quad i=p,n.
\end{equation}
Accurate measurements of neutron skin thicknesses, via future parity violating experiments
\cite{horowitz01} or by means of existing antiprotonic atoms data \cite{brown07,cente09},
are thus helpful in determining the
bulk properties of nuclear systems.

The other important quantity,  the slope of the symmetry energy, is given by
\begin{equation}
L=3 \rho_0 \frac{\partial {\cal E}_{sym}(\rho)}{\partial
  \rho}|_{\rho=\rho_0}, \quad
{\cal E}_{sym} =\left. \frac{1}{2} \frac{\partial^2 {\cal E}/\rho}
{\partial \delta^2} \right|_{\delta=0}.
\end{equation}

\begin{figure}[hbt]
\begin{center}
\begin{tabular}{c}
\includegraphics[width=.9\linewidth]{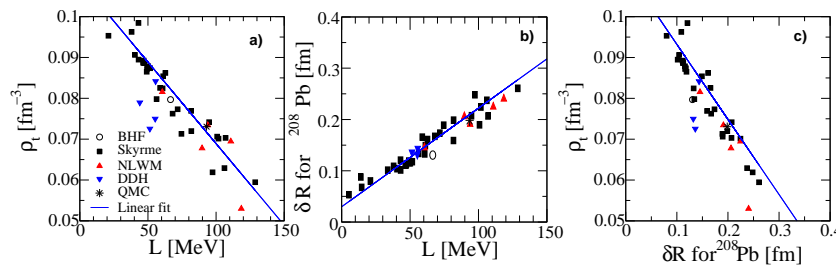}
\end{tabular}
\end{center}
\caption{Correlations between a) the symmetry energy slope $L$ and the transition density
obtained from the thermodynamical spinodal; b) the  symmetry energy slope $L$ and the neutron
skin thickness for $^{208}Pb$; c) the   neutron skin thickness for $^{208}Pb$ and the  the
transition density. Fig. taken from \cite{isaac09}.}
\label{constrain1}
\end{figure}

\begin{figure}[hbt]
\begin{center}
\begin{tabular}{c}
\includegraphics[width=0.4\linewidth]{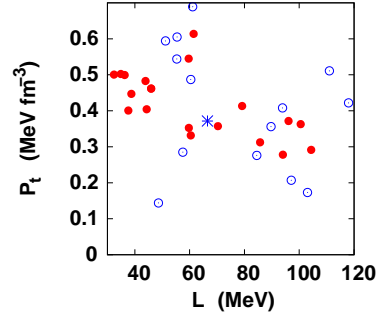}
\end{tabular}
\end{center}
\caption{Transition pressure obtained from the thermodynamical spinodal as a function of the symmetry energy slope $L$ for several Skyrme models (full dots), RMF models (empty dots) and a Brueckner-Hartree-Fock calculation \cite{camille10}.}
\label{constrain2}
\end{figure}

Using a wide range of effective nuclear models, both phenomenological such as the
non-relativistic Skyrme forces or RMF models and microscopic Brueckner-Hartree Fock with the
realistic AV18 potential plus a three-body force of the Urbana type \cite{isaac09},
we have confirmed the
existence of a linear correlation between the symmetry energy slope $L$ and the
crust-core transition density obtained from the thermodynamical spinodal, see Fig. \ref{constrain1}a). The
same models predict  a linear correlation between the neutron skin thickness of the $^{208}$Pb
and $L$ and, consequently, a correlation between  the neutron skin thickness of the $^{208}$Pb
and the transition density $\rho_t$, as proposed in \cite{horo01}. However, in \cite{camille10}
it was shown that no similar correlation exists between the transition pressure and the slope
$L$, Fig. \ref{constrain2},  due to the large dispersion of the predicted
transition  pressure obtained when independent models are considered.
This means that an experimental determination of $L$ alone is not enough for a good estimation of the crust mass
and moment of inertia of a compact star,  since it is the transition pressure that allows a
prediction of the EOS from the observation of glitches \cite{link}.

\section{Diffusion Coefficients and the Pasta Phase}

The neutrino signals detected by astronomers
can be used as a constraint to infer protoneutron star (PNS) composition. For
the same purpose, theoretical studies involving different possible equations of
state obtained for all sorts of matter composition have to be done because the
temporal evolution of the  PNS in the so-called Kelvin-Helmholtz epoch, during
which the remnant compact object changes from a hot and lepton-rich
PNS to a cold and deleptonized neutron star depends on two key ingredients:
the equation of state (EOS) and its associated neutrino opacity
\cite{pons}.  All contributions from neutrino opacities are related with
the diffusion coefficients and can to be used as input
to the solution of the transport equations in the equilibrium
diffusion approximation
to simulate the Kelvin-Helmholtz phase of the protoneutron stars.

The diffusion coefficients that are related to the neutrino opacities are
calculated in such a way that the formation of nuclear pasta at low densities
is taken into account. The diffusion coefficients are given by \cite{pons}
\begin{equation}\label{eq:diff}
D_k=\int_0^{\infty}d\epsilon_\nu~\epsilon_\nu^k \lambda_\nu(\epsilon_\nu)
f_\nu(\epsilon_\nu)(1-f_\nu(\epsilon_\nu)), \quad k=2,3,4,
\end{equation}
where $\lambda_\nu(\epsilon_\nu)$ is the total mean free path of neutrinos and
 $f_{\nu}(\epsilon_\nu)$ is the Fermi-Dirac distribution.

To calculate neutrino opacities and mean free paths we consider neutral current scattering reactions
$ \nu_e+n\rightarrow\nu_e+n$ and $\nu_e+p\rightarrow\nu_e+p$
and charged current absorption reactions
$\nu_e+n\rightarrow e^-+p$ and $\overline{\nu}_e+p\rightarrow e^++n$.
Basically, it consists in calculating the cross sections $\sigma_{n,p}$ for
neutrino-nucleon scattering reactions and the cross section $\sigma_a$ for neutrino absorption
reactions for both nondegenerate and degenerate thermodynamic limits as done in \cite{proto1}.
The thermodynamic regions of intermediate degeneracy are handled by a
simple interpolation. The mean free path is given by
$\lambda_\nu(\epsilon_\nu)=\frac{1}{\rho_n\sigma_n+\rho_p\sigma_p+\rho\sigma_a}$,
where $\rho=\rho_p+\rho_n$ and they are related to the diffusion coefficients by
\begin{equation}
\lambda_\nu^k=\frac{D_k}{\int_0^{\infty}d\epsilon_\nu~\epsilon_\nu^k
f_\nu(\epsilon_\nu) (1-f_\nu(\epsilon_\nu))}.
\end{equation}

The diffusion coefficients $D_2$, $D_3$ and $D_4$ are strongly dependent on
the EOS and are functions of three thermodynamic variables: $\rho$, $T$ and
$Y_L$. We start by fixing $\rho$, $T$ and $Y_L$ from the EOS to calculate the cross
sections $\sigma_p$, $\sigma_n$ and $\sigma_a$ as function of the neutrino
energy and then we integrate in neutrino energy. The numerical
procedure used to calculate the diffusion coefficients
to homogeneous and inhomogeneous matter is the same, except for the nucleon
effective mass. The pasta structure is obtained by the coexistence phases
method, which is based on the enforcement of the Gibbs
conditions. Hence, all other thermodynamic variables (chemical
potentials of all particle species, temperature, pressure and lepton
fraction),  necessary to calculate the neutrino mean free path, and
consequently the diffusion coefficients of the pasta phase are equal in both
phases.  The nucleon effective mass, on the other hand, is not the same.
To calculate $\lambda_\nu(\epsilon_\nu)$ we need  the nucleon effective mass $M^*$.
In the pasta phase, two different phases coexist  (phase I and phase II).
In our calculation, we have used  $M^*=fM^{*(I)}+(1-f)M^{*(II)}$ for the pasta
phase.

\begin{figure}[!htp]
\includegraphics[scale=0.25]{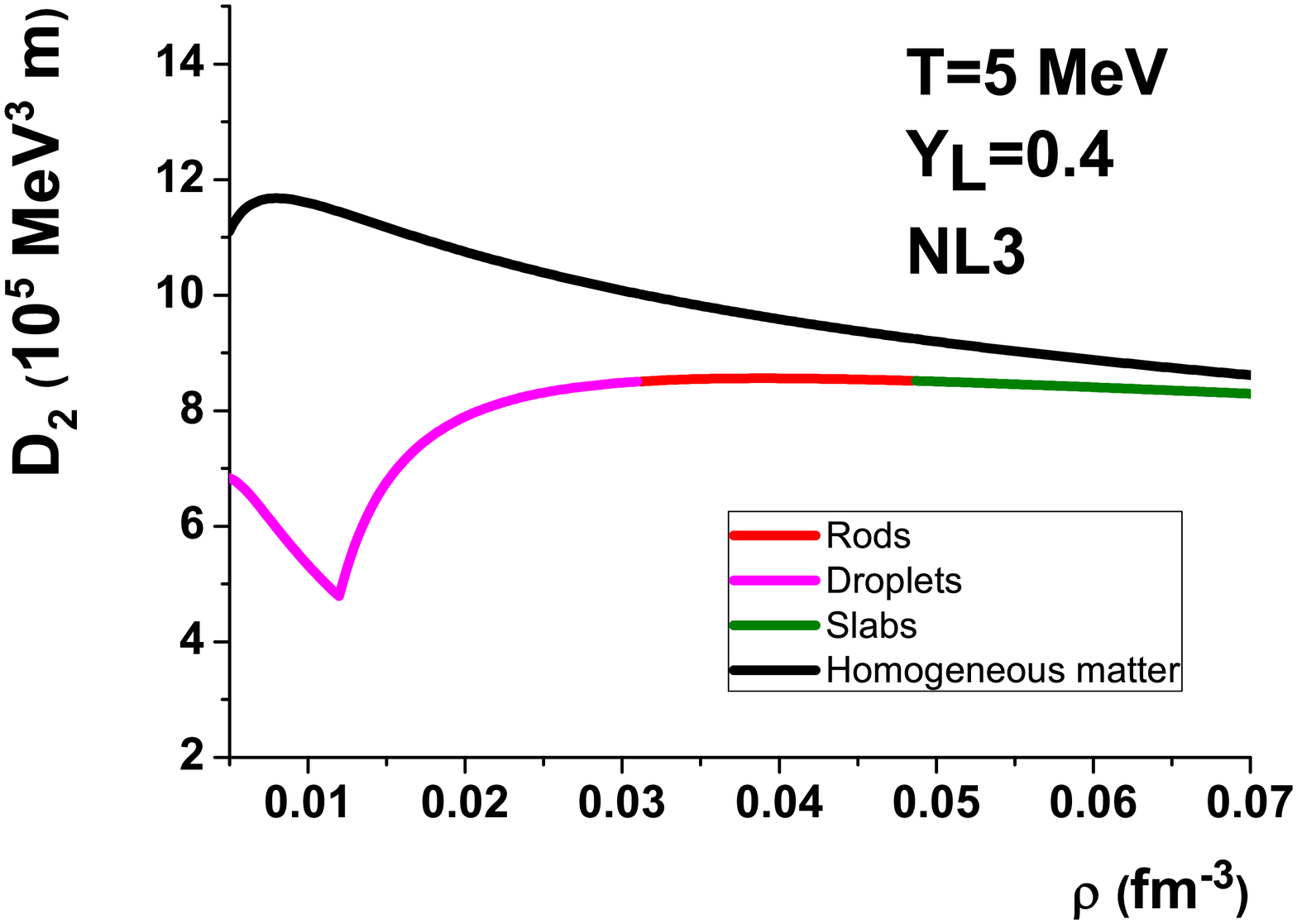}
\includegraphics[scale=0.25]{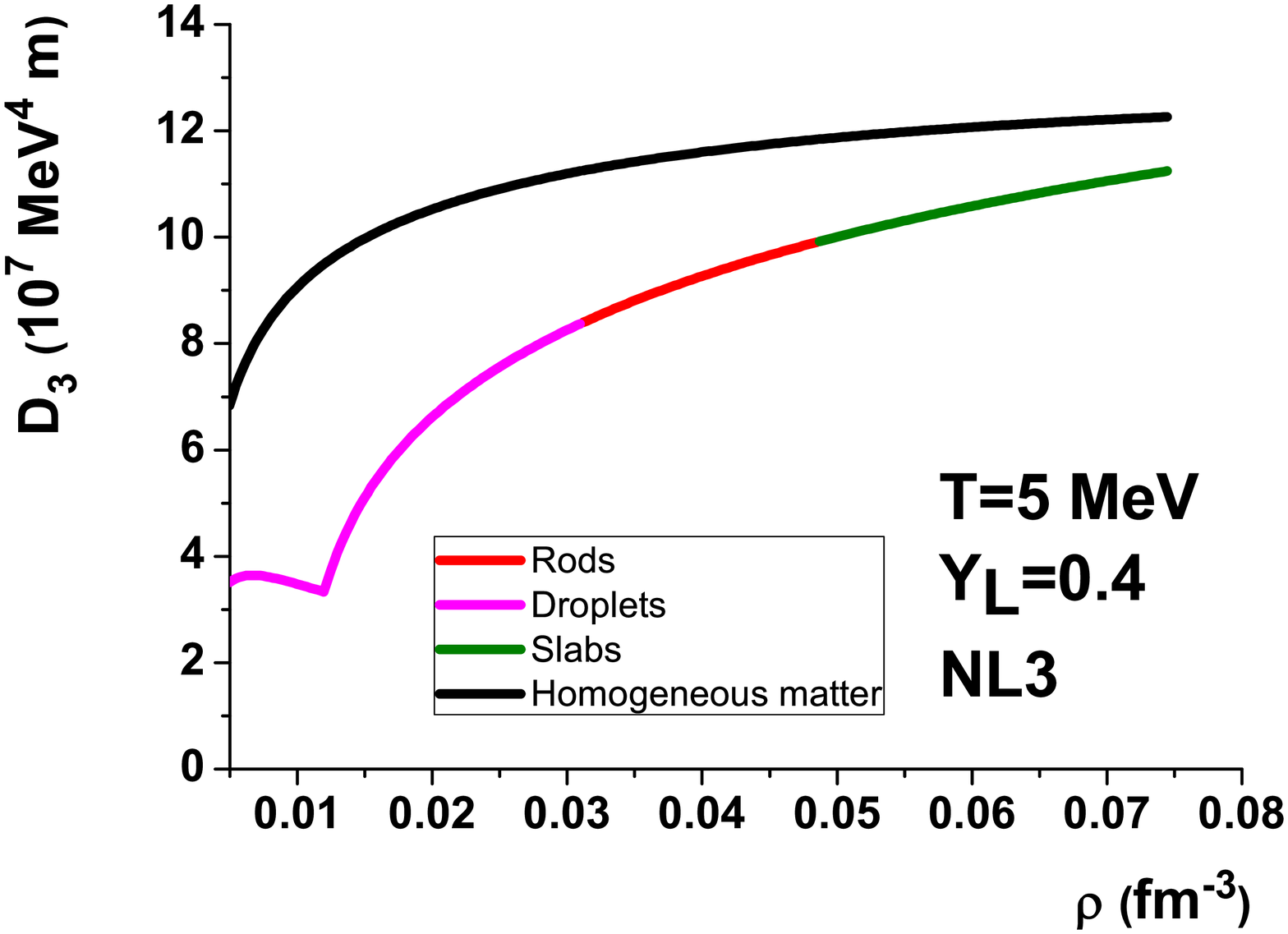}\\
\includegraphics[scale=0.25]{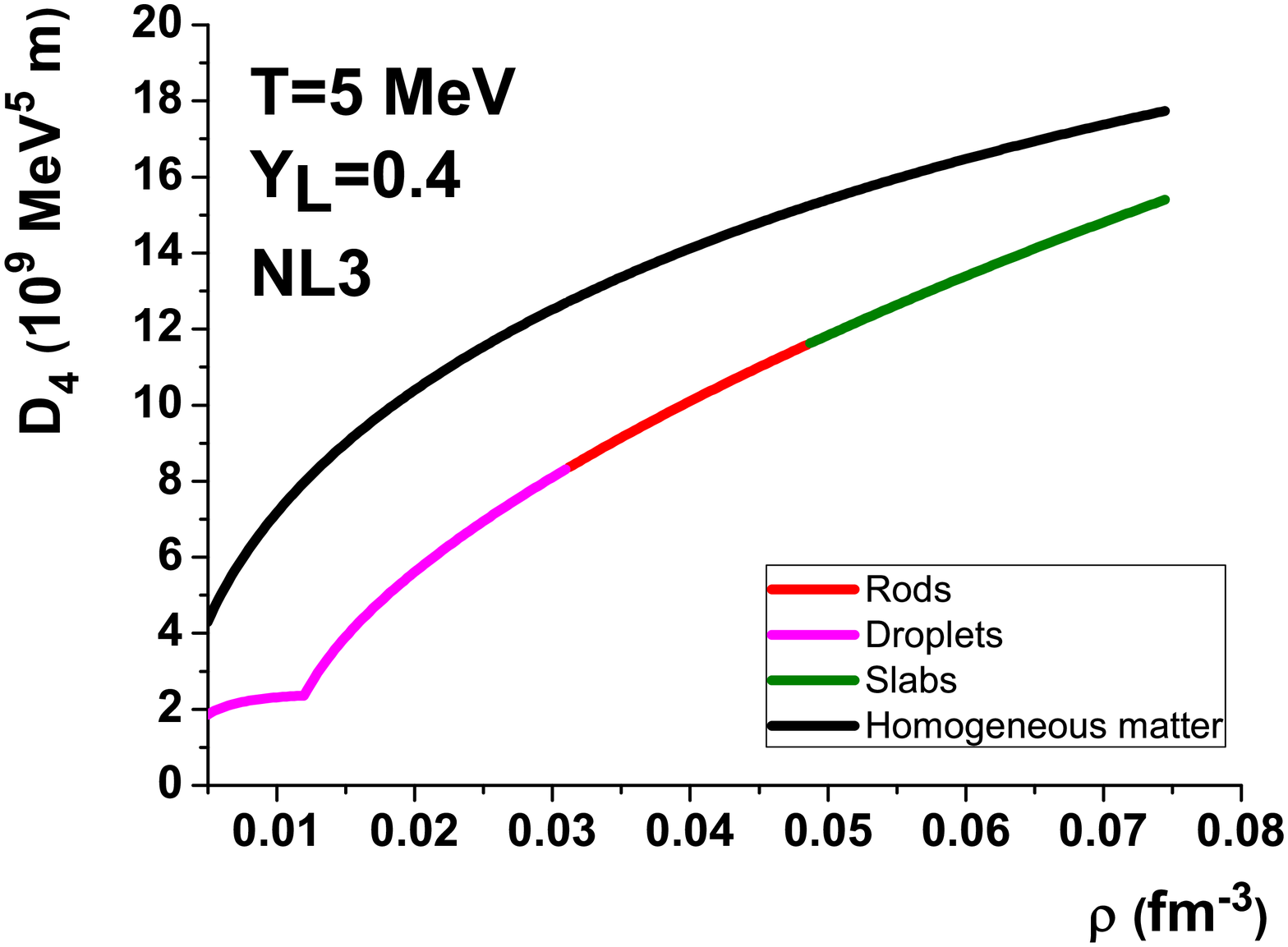}
\caption{\label{dif} Diffusion coefficients $D_2$, $D_3$ and $D_4$ as
function of baryon density at $T=5~MeV$ and $Y_L=0.4$ for homogeneous matter and pasta phase.
Fig. taken from \cite{opacities}.}
\end{figure}

Our results for the diffusion coefficients as a function of the baryon density
at temperature $T=5$~MeV and lepton fraction $Y_L=0.4$ (includes electrons and
trapped neutrinos) is shown in Fig.
\ref{dif}, from where we can see that only three structures are found inside
the pasta phase for the present model: droplets, rods and slabs. While the
diffusion coefficients obtained
with homogeneous matter is always smooth and continuous, a common trend of
all the diffusion coefficients obtained with the pasta phase is a kink at very
low densities in between 0.01 and 0.015 fm$^{-3}$ due to the fact that
the effective nucleon mass becomes greater than the corresponding chemical
potential.

In obtaining the diffusion coefficients, the EOS was calculated as a grid where
temperature ranges are in between 0 and 50 MeV and densities vary from 0.005 to
0.5 fm$^{-3}$.  We have calculated the diffusion coefficients only for baryonic
densities above $0.005~fm^{-3}$ because the integrals of type (\ref{eq:diff})
are very difficult to converge at lower densities.
In all cases the diffusion coefficients obtained with homogeneous matter
join the curves obtained with the pasta phase at densities higher than the
ones shown in Fig \ref{dif}. For $D_2$ calculated at T=5 MeV and $Y_L=0.4$, for instance, they
cross each other at $\rho=0.12$ fm$^{-3}$. Our codes interrupt the calculation
once homogeneous matter becomes the ground state configuration.
This means that there is always a gap in the  diffusion coefficients
when the transport equations are calculated with the
inclusion of the pasta phase.

While the diffusion coefficients obtained with homogeneous matter are
always smooth and continuous, a common trend of
all the diffusion coefficients obtained with the pasta phase is a kink at very
low densities in between 0.01 and 0.015 fm$^{-3}$. The interpolation
procedure we use depends on the quantities $\eta_i=(\mu_i-M^*)/T, i=p,n$.
Whenever either $\eta_p$ or $\eta_n$ inverts its sign, these kinks
appear, i.e., they are the result of the effective nucleon mass being greater
than the corresponding chemical potential.
Moreover, the pasta phase diffusion coefficients  are always
lower than the corresponding coefficients obtained with homogeneous matter.

Our results show that the mean free paths are significantly altered by the
presence of nuclear pasta in stellar
matter when compared with the results obtained with homogeneous matter. These
differences in neutrino opacities
will have consequences in the calculation of the Kelvin-Helmholtz phase of
protoneutron stars \cite{opacities}.

\section{Quark Stars Subject to Strong Magnetic Fields}

Neutron stars with very strong magnetic fields of the order of
$10^{14}-10^{15}$ G are known as magnetars and they are believed to be the
sources of the intense gamma and X rays detected in 1979 \cite{duncan,kouve}.
The hypothesis that some neutron stars are constituted by unbound quark matter
cannot be completely ruled out since the Bodmer-Witten conjecture
\cite{bodmer} cannot be tested in earthly experiments. This conjecture
implies that the true ground state of all matter is (unbound) quark matter
because theoretical predictions show that its energy per baryon at zero
pressure is lower than $^{56}$Fe binding energy. According to \cite{olinto}
the quarks are bound by the strong force rather than the gravitational
force that binds other stars. At the surface of the star, where
the quark density drops abruptly to zero, the electrons extend into a
layer of thickness of the order of $10^3$ fm above the surface, where there
is a super-strong electric field that ties the electrons to the star.
If such a star would be covered by a crust of ordinary nuclear matter
(and neutralizing electrons), it would be blown away as it forms
\cite{Usov1997} or would be destroyed by thermal effects \cite{Kettner1995}.
Thus, one expects quark stars to be bare, in the sense that the surface is
this thin layer of electrons \cite{melrose}.

We investigate quark matter described by the Nambu-Jona-Lasinio \cite{njl}
model exposed to strong  magnetic fields. In the case of pure quark matter,
as predicted by the QCD phase transition possibly taking place in heavy ion collisions,
 the ultra-strong magnetic field results from the superposition of external and internal
fields, the former generated by the alignment of charged particles that are spinning very
rapidly. We use an external field to mimic the real situation.

In order to consider (three flavor) quark matter subject to
strong magnetic fields we introduce the following Lagrangian density
\begin{equation}
{\cal L} = {\cal L}_{f} - \frac {1}{4}F_{\mu \nu}F^{\mu \nu}
\end{equation}
where the quark sector is described by the  su(3) version of the
Nambu--Jona-Lasinio model which includes scalar-pseudoscalar
and the t'Hooft six fermion interaction that
models the axial $U(1)_A$ symmetry breaking:
\begin{equation}
{\cal L}_f = {\bar{\psi}}_f \left[\gamma_\mu\left(i\partial^{\mu}
- q_f A^{\mu} \right)-
{\hat m}_c \right ] \psi_f ~+~ {\cal L}_{sym}~+~{\cal L}_{det}~,
\label{njl}
\end{equation}
where ${\cal L}_{sym}$ and ${\cal L}_{det}$ are given by:
\begin{equation}
{\cal L}_{sym}~=~ G \sum_{a=0}^8 \left [({\bar \psi}_f \lambda_ a \psi_f)^2 + ({\bar \psi}_f i\gamma_5 \lambda_a
 \psi_f)^2 \right ]  ~,
\label{lsym}
\end{equation}
\begin{equation}
{\cal L}_{det}~=~-K \left \{ {\rm det}_f \left [ {\bar \psi}_f(1+\gamma_5) \psi_f \right] +
 {\rm det}_f \left [ {\bar \psi}_f(1-\gamma_5) \psi_f \right] \right \} ~,
\label{ldet}
\end{equation}
where $\psi_f = (u,d,s)^T$ represents a quark field with three
flavors, ${\hat m}_c= {\rm diag}_f (m_u,m_d,m_s)$
 is the corresponding (current) mass matrix while $q_f$
represents the quark electric charge,{  $\lambda_0=\sqrt{2/3}I$  where
$I$ is the unit matrix in the three flavor space, and
$0<\lambda_a\le 8$ denote the Gell-Mann matrices.} Here, we consider
$m_u=m_d \ne m_s$. $A_\mu$ and $F_{\mu \nu }=\partial
_{\mu }A_{\nu }-\partial _{\nu }A_{\mu }$ are used to account
for the external magnetic field. Since we are interested in a
  static and constant magnetic field in the $z$ direction,
$A_\mu=\delta_{\mu 2} x_1 B$. Whenever  $\beta$-equilibrium neutrino-free
  matter is considered, the
leptonic sector is given by eq.(\ref{lage}), where neutrinos are not
taken into account. Besides the su(3) version of the model, we also present two flavor
su(2) NJL model results for comparison.

\begin{figure}[!htp]
\begin{center}
\begin{tabular}{cc}
\includegraphics[width=0.4\linewidth,angle=0]{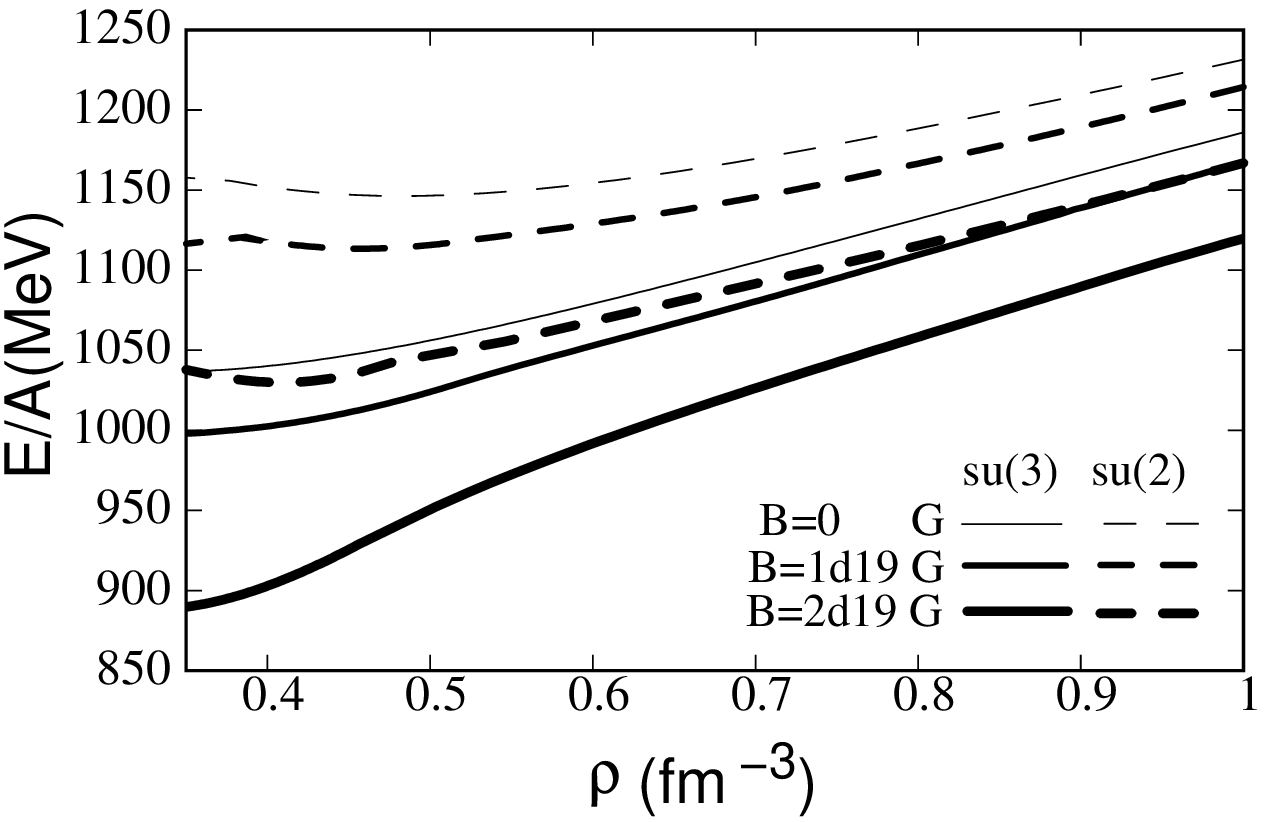} &
\includegraphics[width=0.4\linewidth,angle=0]{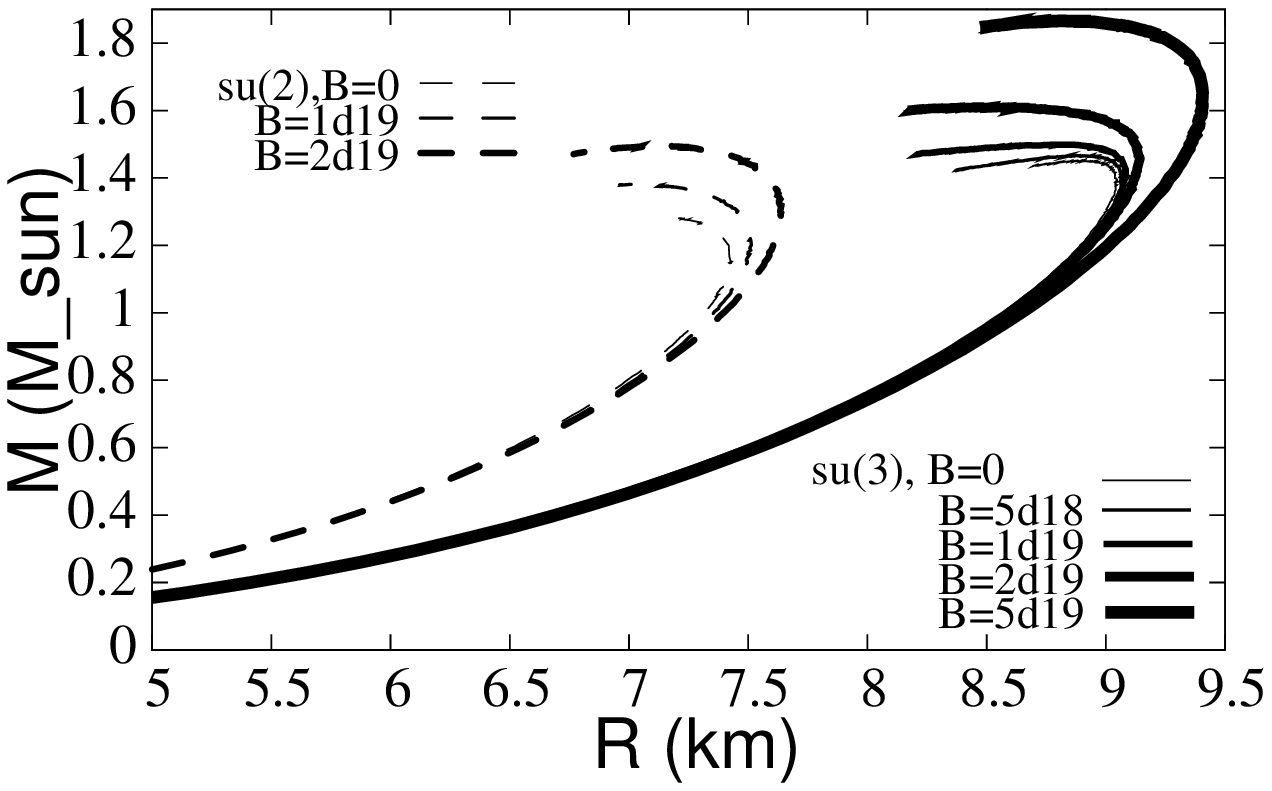}\\
\end{tabular}
\end{center}
\caption{a) Energy per nucleon  as a function of density for $B=0,\, 10^{19}$
and $2\times 10^{19}$ G within NJL $su(2)$ and NJL $su(3)$;
b)Mass-radius curves for the  families of stars within NJL $su(2)$ and NJL
$su(3)$. For $B\le 10^{18}$ the curves coincide with the $B=0$ results.
Fig. taken from \cite{bnjl}.}
\label{ebin}
\end{figure}

In Fig. \ref{ebin}a) one can see that the inclusion of the
magnetic field makes matter more and more bound in both versions of the model.
For the present set of parameters, the energy per baryon $E/A$
of magnetized quark matter becomes more bound than nuclear
matter made of iron nuclei, $\frac{E}{A}|_{^{56}Fe} \sim 930$ MeV
for $B$ around $2 \times 10^{19}$ G.

The EOS for stellar matter under a strong magnetic field is obtained with
  a density dependent frozen magnetic field which is set equal to 10$^{15}$ G
  at the surface and does not exceed $6\times 10^{18}$ G in the center of the star.
Approximate values of the mass and radius of the stars obtained from the integration of the
TOV equations \cite{tov} are displayed
in Fig. \ref{ebin}b), from where it is seen that the
gravitational mass increases with the increase of the
magnetic field for an intensity larger than  $\sim 5 \times 10^{18}$ G for the
su(3) version and $10^{18}$ G for the su(2) NJL.
Another important effect of the field on the properties of the stars is the
increase of the radius of the star. The largest radius may be as
high as 9.5 Km for the $su(3)$ NJL. In general, the maximum mass star
configurations for the  $su(2)$ version of
the NJL model  are smaller with smaller radius, $\sim 7$ Km.

\begin{figure}[!htp]
\begin{center}
\begin{tabular}{cc}
\includegraphics[width=0.4\linewidth]{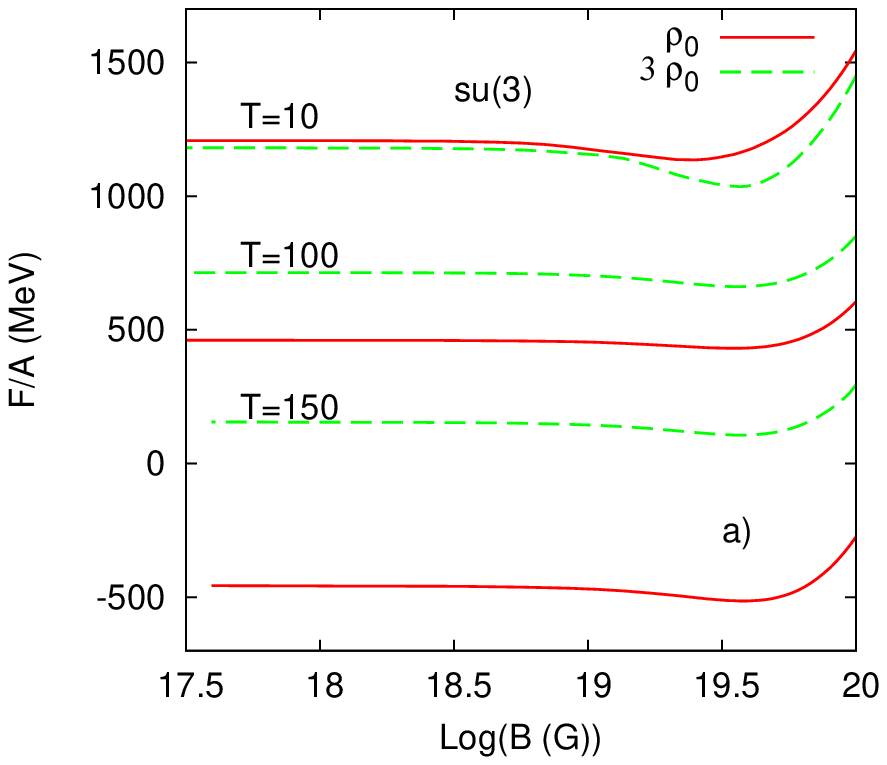} &
\includegraphics[width=0.4\linewidth]{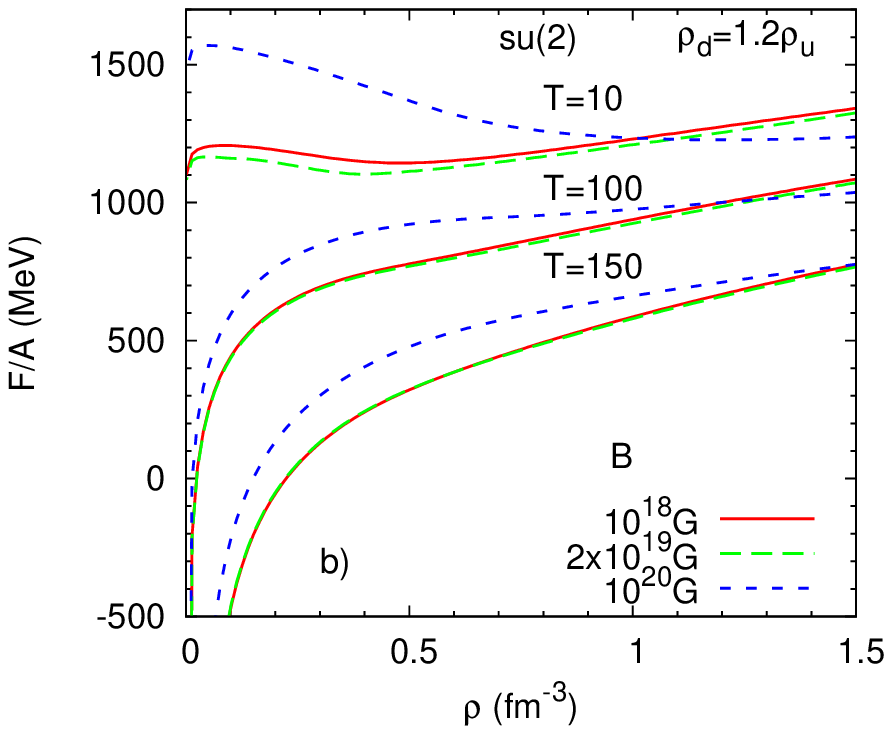} \\
\end{tabular}
\end{center}
\caption{Free energy per particle for NJL a) su(3)  at  $T=10,\, 100,\, 150$ MeV
 as a function of B  for symmetric quark matter and  $\rho=\, \rho_0$ and
$3\rho_0$; b) su(2) versus the density for asymmetric matter
$\rho_d=1.2 \rho_u$, $T=10,\, 100,\,150$ MeV and $B=10^{18}, 2\times 10^{19},
\, 10^{20}$ G. Fig. taken from \cite{tnjl}}
\label{tnjl}
\end{figure}

Next we discuss the effect of the magnetic field and temperature on
quark matter as possibly formed in heavy-ion reactions. We  display the free energy per
particle in terms of the magnetic field for symmetric matter in Fig. \ref{tnjl}a)  and in terms of
the density for asymmetric matter in Fig.\ref{tnjl}b). As discussed before,
the effect of the magnetic field is stronger for smaller densities and
temperatures. At a given density the main effect of temperature is to
decrease the free energy per particle.
The effect of the magnetic field is clearly seen in Fig. \ref{tnjl}a). It is
stronger for the smaller temperatures and, for a strong enough field it gives rise to a decrease of the free energy (above $B=4\times 10^{18}$ G
for $\rho=\rho_0$), due to a reduction of the number of Landau levels. However, for even
stronger fields, the free energy increases due to an increase of the
effective quark masses with $B$. This explains why in  Fig.\ref{tnjl}b) the free energy
at T=10 MeV below $\rho=0.5$ fm$^{-3}$ is smaller for $B=2\times 10^{19}$ G  and
larger for $B=10^{20}$ G. For larger temperatures, the  reduction observed at
intermediate densities
washes out and it is only observed an increase of the free energy for fields
above $~5\times 10^{19}$ G.
 The increase of the free energy for very large fields is mainly due to a
reduction of  the entropy and an increase of
the effective mass  for  $B>~10^{19}$ G.

To extend the present calculation in order to consider matter
 in protoneutron stars, the enforcement of charge neutrality
  and $\beta$-equilibrium is required. The ideal calculation would
  consider a fixed entropy resulting in temperatures lower than 50 MeV
  in the interior of the stars. This extension is simple and straightforward.

\section{Conclusions and open questions}

In this chapter we have focused on the study of possible internal structures
of the neutron stars crust. The existence of the pasta phase is
related to the instabilities and possible coexisting phases in the
nuclear matter, features well described by the spinodal and binodal sections.
Important remaining questions are: at which temperatures do the pasta
structures dissolve? Up to which temperatures is the
ansatz of a Wigner Seitz cell valid? We have tried to validate a
simple coexistence phase approach to obtain the pasta phase and its
structure by using the results of possible surface energy parametrizations
and comparing the final results with the more realistic and self-consistent Thomas-Fermi
approximation, which is very time consuming. It remains to be checked
how the Thomas-Fermi calculation of $\beta$-equilibrium pasta phase
compares with the results obtained within Hartree-Fock and
Hartree-Fock-Bogoliubov, where only spherical clusters are considered. More
calculations in this direction are still required.

So far, we have included $\alpha$ particles as part of the pasta phase.
Will light clusters other than $\alpha$ particles contribute in a significative
manner to the extension of the pasta phase and its transition density to
homogeneous matter? Moreover, how will the presence of light clusters
affect the properties of the pasta phase? Are the values
for the dissolution density of the different light clusters in nuclear
matter realistic \cite{pasta_clusters}?

We have also seen that the inclusion of the pasta phase affects the
neutrino opacities through its diffusion coefficients. It remains to be
checked how the internal structure of the pasta phase, which is model
and approximation dependent, influences the diffusion coefficients
through the different neutrino mean free paths. A more complete
calculation will show us how the existence of the  pasta
structures affect the transport properties of the crust.

Our last section was devoted to a subject other than the crust of
neutron stars. The measurement of a pulsar of inferred mass
$1.97 \pm 0.04 M_\odot$ known as PSR J1614-2230 \cite{ozel}, opened new
questions about its possible constituents and related equation of state. It is well known that there
is a maximum mass that can be supported against collapse and its theoretical
value is very model dependent. According to \cite{kurkela}, the maximum mass of a quark
star lies between 2 and 2.7 $M_\odot$. Hence, it is claimed that  PSR J1614-2230 can be a quark
star. Do quark stars really exist? How are they affected by strong magnetic
fields?  We have seen that our results for the NJL model cannot reproduce
such high masses, not even with very strong magnetic fields.

%\vspace{0.5cm}

{\bf Acknowledgments} - This work was partially supported by the
Capes/FCT n. 232/09 bilateral collaboration, by CNPq and FAPESC (Brazil),
by FCT and COMPETE/FEDER (Portugal) and  by Compstar, an ESF Research
Networking Programme. We thank Drs. M. Benghi, M.M. W. de Moraes,
J. R. Marinelli, C.C. Barros, L. Brito, S. Chiacchiera, A.M.S. Santos,
I. Vida\~na, C. Ducoin, Ph. Chomaz, D.B. Melrose, A.P. Martinez and A. Rabhi
for our long term collaborations on some of the topics discussed
in this
chapter.

\label{lastpage-01}


\begin{thebibliography}{99}

\bibitem{glen} N. K. Glendenning, Compact Stars, Springer-Verlag, New-York,
2000.

\bibitem{bodmer} N. Itoh, \textit{Prog. Theor. Phys.} 44, { 291} (\textbf{1970});
A.R. Bodmer, \textit{Phys. Rev. D} {  4}, 1601 (\textbf{1971});
E. Witten, \textit{Phys. Rev. D} {  30}, 272 (\textbf{1984}).
\bibitem{olinto} C. Alcock, E. Farhi and A. Olinto, \textit{Astrophys. J.} { 310}, 261 (\textbf{1986}).

\bibitem{tw} S. Typel and H. H. Wolter, \textit{Nucl. Phys. A} { 656}, 331 (\textbf{1999});
Guo Hua, Liu Bo and M. Di Toro, \textit{Phys. Rev. C }{   62}, 035203(\textbf{2000}).

\bibitem{haensel} P. Haensel,  Final Stages of Stellar Evolution, ed. J.-M. Hameury and C. Motch,  (EDP Sciences,
2003), EAS Publ.Ser.  { 7} (2003) 249.

\bibitem{gaitanos} T. Gaitanos, M. Di Toro, S. Typel, V. Baran, C. Fuchs,
V. Greco  and H. H. Wolter, \textit{Nucl. Phys. A} {  732}, 24 (\textbf{2004}).

\bibitem{inst04} S.S. Avancini, L. Brito, D. P. Menezes and C. Provid\^encia,
\textit{Phys. Rev. C} { 70}, 015203 (\textbf{2004}).

\bibitem{nl3} G. A. Lalazissis, J. K\"onig and P. Ring, \textit{Phys. Rev. C} {  55},
540 (\textbf{1997}).

\bibitem{pethick95}C. J. Pethick and D. G. Ravenhall, \textit{Annu. Rev. Nucl. Part. Sci.} 45, 429 (\textbf{1995}).
\textit{Phys. A} {  584}, 675 (\textbf{1995}).

\bibitem{link} B. Link, R.I. Epstein, J.M. Lattimer, \textit{Phys. Rev. Lett.} { 83} (\textbf{1999}) 3362.

\bibitem{muller95} H. M\"uller and B. D. Serot, \textit{Phys. Rev. C} {  52}, 2072
(\textbf{1995}).

\bibitem{avancini06} S. S. Avancini, L. Brito, Ph.
Chomaz, D. P. Menezes, and C. Provid\^encia,\textit{ Phys. Rev. C} {  74}, 024317 (\textbf{2006}).

\bibitem{pethick95a} C. J. Pethick, D. G. Ravenhall, and C. P. Lorenz, \textit{Nucl. Phys. A} { 584}, 675 (\textbf{1995}).

\bibitem{brito06} C. Provid\^encia, L. Brito, S. S. Avancini, D. P. Menezes, and Ph. Chomaz,
\textit{Phys. Rev. C} { 73}, 025805 (\textbf{2006}); L. Brito, C. Provid\^encia, A. M. Santos, S. S. Avancini, D. P. Menezes, and Ph. Chomaz, \textit{Phys. Rev. C} { 74}, 045801 (\textbf{2006}).

\bibitem{erratum} S.S. Avancini, S. Chiacchiera, D.P. Menezes and C.
Provid\^encia, arXiv:1010.3644v3[nucl-th], \textit{ Phys. Rev. C}
(\textbf{2012}), in press.

\bibitem{warmpasta} S.S. Avancini, S. Chiacchiera, D.P. Menezes and C.
Provid\^encia,\textit{ Phys. Rev. C} {  82}, 055807 (\textbf{2010}).

\bibitem{pasta_alpha} S.S. Avancini, C.C. Barros, D.P. Menezes and C.
Provid\^encia,\textit{ Phys. Rev. C} {  82}, 025808 (\textbf{2010}).

\bibitem{watanabe08} H. Sonoda, G. Watanabe, K. Sato, K. Yasuoka, and T. Ebisuzaki, \textit{Phys. Rev. C} { 77}, 035806 (2008); { ibid}, \textit{Phys. Rev. C} { 81}, 049902 (\textbf{2010}).

\bibitem{newton09} W. G. Newton and J. R. Stone, \textit{Phys. Rev. C} { 79}, 055801 (\textbf{2009}).

\bibitem{blaschke09} S. Typel, G. Roepke, T. Klahn, D. Blaschke and H.H. Wolter,
\textit{Phys. Rev. }{ C 81}, 015803 (\textbf{2010}).


\bibitem{li08} B. A. Li, L. W. Chen and C. M. Ko, \textit{Phys. Rep.} { 464}, 113 (\textbf{2008}).
\bibitem{garg07} U. Garg {\it et al.,}\textit{ Nucl. Phys. A} { 788}, 36 (\textbf{2007}).
\bibitem{danie09} P. Danielewicz and J. Lee, \textit{Nucl. Phys. A} { 818}, 36 (\textbf{2009}).
\bibitem{li05} B. A. Li, G.-C. Yang and W. Zuo, \textit{Phys. Rev. C} { 71}, 014608 (\textbf{2005}).
\bibitem{fuchs06} C. Fuchs, Prog. Part. \textit{Nucl. Phys.} { 56}, 1 (\textbf{2006}).
\bibitem{yennello10} D.V. Shetty, S.J. Yennello,  Pramana, { 75} (2010) 259; arXiv:1002.0313v4 [nucl-ex].
\bibitem{brown00} B. A. Brown, \textit{Phys. Rev. Lett.} { 85}, 5296 (\textbf{2000}).

\bibitem{typel01} S. Typel and B. A. Brown, \textit{Phys. Rev. C} { 64}, 027302 (\textbf{2001}).
\bibitem{horo01} C. J. Horowitcz and J. Piekarewicz,\textit{ Phys. Rev. Lett.} { 86}, 5647 (\textbf{2001}).
\bibitem{horowitz01} C. J. Horowitz, S. J. Pollock, P. A. Souder and R. Michaels,  \textit{Phys. Rev. C }{ 63}, 025501 (\textbf{2001}).

\bibitem{brown07} B. A. Brown, G. Shen, G. C. Hillhouse, J. Meng and A. Trzcinska, \textit{Phys. Rev. C} { 76}, 034305 (\textbf{2007}).
\bibitem{cente09} M. Centelles, X. Roca-Maza, X. Vi\~nas and M. Warda, \textit{Phys. Rev. Lett.} { 102}, 122502 (\textbf{2009}).

\bibitem{isaac09} I. Vida\~na, C.  Provid\^encia, A. Polls, and
  A. Rios,\textit{ Phys. Rev. C} { 80}, 045806 (\textbf{2009})

\bibitem{camille10} C. Ducoin, J. Margueron, and C. Provid\^encia, \textit{Europhys. Lett.} { 91}, 32001 (\textbf{2010}).

\bibitem{pons} J.A. Pons, A.W. Steiner, M. Prakash and J.M. Lattimer,
\textit{Phys. Rev. Lett.} { 86} (\textbf{2001}) 5223-5226; J.A. Pons, S. Reddy, M. Prakash,
J.M. Lattimer and J.A. Miralles, \textit{Astrophys. J.} { 513}, 780 (\textbf{1999});
S. Reddy, M. Prakash and J.M. Lattimer, \textit{Phys. Rev. D }{  58}, 013009 (\textbf{1998}).

\bibitem{proto1} Burrows, A. and Lattimer, J.~M., \textit{Apj}, { 307}, 178,(\textbf{1986}).

\bibitem{opacities} M.D. Alloy and D.P. Menezes, \textit{Phys. Rev. C} {
    83}, 035803 (\textbf{2011}).

\bibitem{duncan} R. Duncan and C. Thompson, \textit{Astron. J} { 32}, L9 (\textbf{1992});
astro-ph/0002442.

\bibitem{kouve} C. Kouveliotou et al, \textit{Nature} { 393}, 235 (\textbf{1998}).

\bibitem{Usov1997} V.V. Usov, \textit{ApJ} { 481}, L107 (\textbf{1997}).

\bibitem{Kettner1995}C. Kettner, F. Weber, M.K. Weigel, N.K. Gledenning,
\textit{Phys. Rev. D} {  51}, 1440 (\textbf{1995}).

\bibitem{melrose} D.B. Melrose, R. Fock and D.P. Menezes, \textit{Month. Not. Roy.
Astr. Soc.} { 371}, 204 (\textbf{2006}).

\bibitem{njl} Y. Nambu and G. Jona-Lasinio, \textit{Phys. Rev.} {  122}, 345 (\textbf{1961});
{ 124}, 246 (\textbf{1961}).

\bibitem{bnjl} D.P. Menezes, M. Benghi Pinto, S.S. Avancini and C.
Provid\^encia,\textit{ Phys. Rev. C }{  80}, 065805 (\textbf{2009}).

\bibitem{tov} Tolman, R.C., \textit{Phys. Rev.} { 55} (\textbf{1939}) 364; J.R. Oppenheimer and
G.M. Volkoff, \textit{Phys. Rev.} { 55} (\textbf{1939}) 374.

\bibitem{tnjl} S.S.Avancini, D.P. Menezes and C. Provid\^encia,
 \textit{ Phys. Rev. C }{  83}, 065805 (\textbf{2011}) .

\bibitem{pasta_clusters} S.S. Avancini, C.C. Barros, L. Brito, S. Chiacchiera, D.P. 
Menezes and C. Provid\^encia, \textit{Phys. Rev. C}{85}, 035806 (\textbf{2012}).

\bibitem{ozel} F. Ozel, D. Psaltis, S. Ransom, P. Demorest and M. Alford,
\textit{ApJ} { 724}, L199 (\textbf{2010}).

\bibitem{kurkela} A. Kurkela, P. Romatschke and A. Vuorinen, \textit{Phys. Rev.
D} {  81}, 105021 (\textbf{2010}).


\end{thebibliography}
\end{document}